\documentclass[prl,aps,twocolumn,showpacs ,10pt]{revtex4-1}
\usepackage{amsmath,amssymb,graphicx}
\usepackage{epsfig}
\usepackage{textcomp}
\usepackage{color}
\usepackage{epstopdf}
\usepackage{array}
\usepackage[normalem]{ulem}
\usepackage{graphicx}

\usepackage{mathrsfs}
\usepackage[utf8]{inputenc}
\usepackage[english]{babel}
\usepackage{amsthm}

\usepackage{mathrsfs}
\usepackage[utf8]{inputenc}
\usepackage[english]{babel}
\usepackage{amsthm}

\usepackage{bm}

\newcommand{\OO}{\mathcal{O}}

\newcommand{\matarrow}[1]{\overset{\text{\tiny$\leftrightarrow$}}{#1}}

\begin{document}
\title{Extinction risk of a Metapopulation under the Allee Effect}
\author{Ohad Vilk and Michael Assaf}
\email{michael.assaf@mail.huji.ac.il}

\affiliation{Racah Institute of Physics, Hebrew University of Jerusalem, Jerusalem 91904, Israel}

\begin{abstract}
We study the extinction risk of a fragmented population residing on a network of patches coupled
by migration, where the local patch dynamics include the Allee effect. We show that mixing between
patches dramatically influences the population's viability. Slow migration is
shown to always increase the population's global extinction risk compared to the isolated case. At fast migration, we demonstrate that synchrony
between patches minimizes the population's extinction risk. Moreover, we discover a critical migration rate that maximizes the extinction risk of the population, and identify an early-warning signal when approaching this state. Our theoretical results are confirmed via the highly-efficient weighted ensemble method. Notably, our analysis can also be applied to studying switching in gene regulatory networks with multiple transcriptional states.
\end{abstract}
\maketitle

Extinction of a metapopulation -- a network of interacting spatially-separated populations (patches) of the same species -- is of key interest in various scientific disciplines such as ecology, evolutionary biology and genetics~\cite{hanski2004ecology}. Here a major challenge is finding the optimal interaction strategy among the individual patches in such a fragmented population, that maximizes the metapopulation lifetime. Under certain conditions, it has been found that interactions between the patches decrease the extinction risk of the metapopulation, while in other cases the opposite may occur, \textit{i.e.}, isolation of the individual patches minimizes the population's extinction risk~\cite{fahrig2017ecological,wilcox1985conservation}.

In previous studies, metapopulation dynamics have mostly been modelled at the deterministic level, using various versions of the Levins model~\cite{levins1969some,hanski2004ecology}. Several of these models incorporated the so-called Allee effect~\footnote{The Allee effect is a phenomenon in population biology which gives rise to a negative per-capita growth rate at small population
sizes, yielding a critical population density, or colonization threshold, under which extinction occurs deterministically~\cite{stephens1999allee}.}, but only at the metapopulation level~\cite{lande1987extinction,amarasekare1998allee, lande1998extinction, hanski2000metapopulation, ovaskainen2001spatially}. In recent studies, demographic noise, stemming from the discreteness of individuals and stochasticity of the birth-death interactions, has also been accounted for allowing \textit{e.g.} the calculation of the mean time to extinction (MTE) of the metapopulation~\cite{khasin2012minimizing,khasin2012fast,eriksson2014emergence, assaf2017wkb, ovaskainen2017interplay}. However, non of these works has conducted a systematic study on the extinction risk of a metapopulation, while incorporating the Allee effect at the individual patch level.

In many realistic examples of metapopulations, it has been shown that the Allee effect is present and plays a crucial role in the dynamics of the population~\cite{kuussaari1998allee, hanski2004ecology,kramer2009evidence}. Furthermore, at the level of the individual patch, incorporating the Allee effect can have important consequences on the population's extinction risk, and thus affects population management and preservation~\cite{dennis2002allee, taylor2005allee}. As a result, it is obvious that within a metapopulation, the Allee effect can strongly influence both extinction and colonization of individual patches, as well as global metapopulation extinction~\cite{hanski2004ecology,khasin2012fast}, and therefore it is vital to take the Allee effect into account when dealing with metapopulation extinction.

In this manuscript we reveal novel metapopulation behavior when the local birth-death dynamics on each patch exhibit the Allee effect, by coupling the local demographic noise to stochastic migration between patches. When migration is slow, we show that the system displays multiple stable fixed points (FPs) at the deterministic level, which become metastable at the stochastic level, giving rise to the existence of multiple routes to extinction. For fast migration, synchrony drives the population to a maximum of two stable states at the deterministic level, and at the stochastic level we find that the extinction risk is minimized when the typical flux across patches is comparable. Importantly, we demonstrate the exact conditions for which mixing (at some migration rate) or complete isolation, is optimal for the population's mean lifetime. Our theoretical analysis relies on the Wentzel–Kramers–Brillouin (WKB) approximation at the master equation level. The resulting Hamiltonian can be analytically dealt with in the limit of slow and fast migration.
Moreover, our results drastically simplify close to bifurcation, where the colonized and colonization threshold states, merge (see below).
Our theoretical results are tested against highly-efficient numerical simulations based on the weighted-ensemble (WE) method.

\textit{Model and deterministic dynamics.} We consider $M$ patches, where migration between patch $i$ and $j$ occurs at a rate $\mu_{ij}$. To locally account for the Allee effect on each patch $i=1,\dots,M$, we chose a simple birth-death process, $2A \leftrightarrow 3A$ and $A\rightarrow 0$, that gives rise to bistable dynamics~\cite{assaf2017wkb}, such that on each patch we have a single-step birth-death process~\cite{khasin2012fast}:
\begin{eqnarray}\label{rates}
&&n_i \xrightarrow{B_{n_i}}n_i+1,\;\;\;\;\;\;\;\;\;\;\; n_i \xrightarrow{D_{n_i}}n_i-1\\
&&B_{n_i}= \frac{2n_i(n_i -1)}{N_i (1-\delta_i^2)},\;\;\;\; D_{n_i}= n_i + \frac{n_i(n_i -1)(n_i -2)}{N_i^2 (1-\delta_i^2)}\nonumber.
\end{eqnarray}
Here, $0<\delta_i < 1$ determines the distance between the local colonized and colonization threshold states, while $N_i\gg 1$ is the local carrying capacity, see below. Denoting $\kappa_i=N_i/N_1$, the deterministic rate equation for the population density at patch $i$, $x_i=n_i/N_1$, reads
\begin{equation} \label{RateEqMultiPatch_x}
\dot{x}_i=  b_i(x_i) - d_i(x_i) + \sum_{j \neq i}^M (\mu_{ji} x_j -  \mu_{ij} x_i).
\end{equation}
Here, $b_i(x_i)=B_{n_i}/N_1$ and $d_i(x_i) = D_{n_i}/N_1$, see Eqs.~(\ref{rates}), while ${\cal O}(N^{-1})$ terms were neglected.

For an isolated patch $i$, Eq.~(\ref{RateEqMultiPatch_x}) has three FPs: $\bar{x}_{i,0} = 0$ and $\bar{x}_{i,+} = \kappa_i(1+\delta_i)$ are stable, corresponding to the extinct and colonized states, respectively, while $\bar{x}_{i,-} =  \kappa_i(1 - \delta_i)$ is unstable, and corresponds to the colonization threshold~\cite{stephens1999allee}. Note that the relaxation time in the vicinity of the colonized state, $t_{i}^{relax} = (1 - \delta_i) / (2 \delta_i)$, determines the typical time scale of the local patch dynamics, where in general we have $t_{i}^{relax}={\cal O}(1)$. Finally, Eq.~(\ref{RateEqMultiPatch_x}) is valid as long as $N_i(1 - \delta_i) \gg 1$, and $N_i\mu_{ij} \gg 1$.

\begin{figure}[t]
	\includegraphics[width=0.95\linewidth]{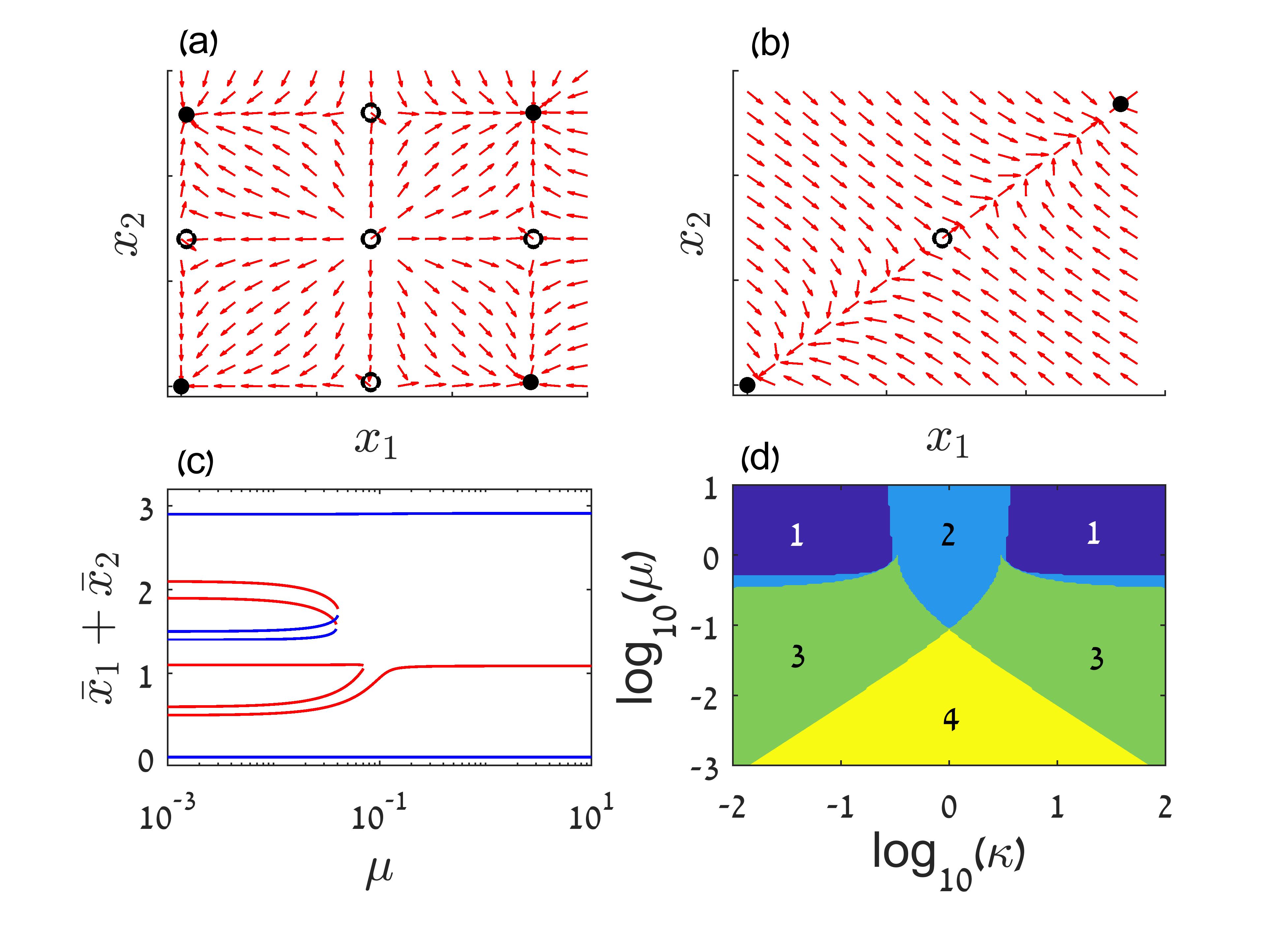}
\vspace{-2mm}
	\caption{(a) and (b): Dynamical trajectories of Eqs.~(\ref{RateEqMultiPatch_x}) in the case of two patches, with  $\kappa = 1$, $\alpha = 1$, $\delta_1 = \delta_2 = 0.3$ and (a) $\mu = 0.01$, (b) $\mu = 1$. Stable and unstable FP are denoted by full and open circles, respectively.
(c): Bifurcation diagram of Eqs.~(\ref{RateEqMultiPatch_x}), with $\kappa = 1$, $\alpha = 1$, $\delta_1 = 0.5$, $\delta_2 = 0.4$, as a function of $\mu$. Here, the blue and red lines correspond to stable and unstable FPs, respectively. (d): Number of stable FPs as function of both $\kappa$ and $\mu$, for $\alpha = 1$, $\delta_1 = \delta_2 = 0.5$.}
	\label{fig1}
\end{figure}

While most of the analysis below can be generalized for $M$ patches~\cite{SM}, here we focus on the dynamics of two connected patches, which capture most of the interesting features in this problem. Thus, henceforth we have $M=2$ and denote $\mu_{12} \equiv \mu$ and $\mu_{21} \equiv \alpha \mu$, where $\alpha={\cal O}(1)$ is the ratio between the migration rates, while $\kappa_1=1$ and $\kappa_2 \equiv \kappa={\cal O}(1)$ is the carrying capacities ratio.

At the deterministic level, the dynamics for slow and fast migration is markedly different. For slow migration, $\mu\ll 1$, \textit{i.e.}, when the typical time scale of migration is slow compared to that of the local patch, Eqs.~(\ref{RateEqMultiPatch_x}) give rise to a maximum of nine FPs, four of which are stable, see Fig.~\ref{fig1}(a,c,d). Here, the FPs are shifted by $\mathcal{O}(\mu)$ compared to the case where the patches are isolated, see~\cite{SM} for detailed calculations and further examples. For fast migration, $\mu\gg 1$, there are at most three FPs, two of which are stable, see Fig.~\ref{fig1}(b-d). In this case, in the leading order in $\mu\gg 1$, patches are synchronized and the FPs of Eqs.~(\ref{RateEqMultiPatch_x}) satisfy $\bar{x}_2 = \bar{x}_1/\alpha $, with $\bar{x}_{1,0} = 0$ and $\bar{x}_{1,\pm} = \tilde{\kappa}(1\pm \tilde{\delta})$, such that the combined size of the colonized patches is $(1+1/\alpha)\bar{x}_{1,+}$. Here, $\tilde{\kappa}$ and $\tilde{\delta}$ are the effective carrying capacity and threshold parameters, respectively, and we demand that $0<\tilde{\delta}<1$ is real; otherwise, deterministic extinction occurs~\cite{SM}. In Fig.~\ref{fig1}(c) by numerically solving Eqs.~(\ref{RateEqMultiPatch_x}) for the entire range of $\mu$, we demonstrate the multiple bifurcations occurring as $\mu$ increases. Finally, in Fig.~\ref{fig1}(d) we map the number of stable FPs as a function of both $\mu$ and $\kappa$ displaying a reduction in the number FPs as $\mu$ increases, and as $\kappa$ diverges from $1/\alpha$, see below. Additional examples of the deterministic dynamics can be seen in Fig.~S1.

\textit{Stochastic formulation.} To account for local demographic noise and the stochastic migration across patches, we write down the master equation describing the evolution of $\mathbb{P}_{n_1,n_2}$ -- the probability of finding $n_1$ and $n_2$ individuals in patch 1 and 2, respectively, at time t:
\begin{eqnarray} \label{MasterEquationDdim}
&&\dot{\mathbb{P}}_{n_1,n_2}= \left\{ \sum_{i = 1}^2  \left[(\mathbb{E}_{n_i}^{-1} - 1) B_{n_i} + (\mathbb{E}_{n_i}^{1} - 1)D_{n_i}\right] \right.  \\
&& \left.+ \left[(\mathbb{E}_{n_1}^{1}\mathbb{E}_{n_2}^{-1} - 1)n_1 + (\mathbb{E}_{n_1}^{-1}\mathbb{E}_{n_2}^{1}  - 1)\alpha n_2 \right] \mu \right\} \mathbb{P}_{n_1,n_2}, \nonumber
\end{eqnarray}
where $\mathbb{E}_{n_i}^{\pm 1} f(n)=f(n \pm 1)$.
In the absence of external flux into either of the patches, starting from any initial condition, the system ultimately undergoes extinction, where $\mathbb{P}_{0,0}$ grows in time while all other probabilities decay. Yet, in the limit of large carrying capacities, the decay rate turns out to be exponentially small, see below, and one can use the metastable ansatz $\mathbb{P}_{n_1,n_2}=\pi_{n_1,n_2} \exp(-t/\tau)$, where $\pi_{n_1,n_2}$ is the quasi-stationary distribution and $\tau$ is the MTE. The latters can be found by employing the WKB ansatz $\pi_{n_1,n_2} \sim \exp[-N S(x_1,x_2)]$, where $S$ is the action function~\cite{Dykman1994}, arriving at a stationary Hamilton-Jacobi equation, $H=0$~\cite{Dykman1994, assaf2006spectral, kessler2007extinction, meerson2008noise,escudero2009switching, assaf2010extinction}, with Hamiltonian
\begin{eqnarray} \label{HamiltonianDdim}
H(x_1,p_1,x_2,p_2)&=&\sum_{i =1}^2 \left(e^{p_{i}}\!-\!1\right) \left[b_i(x_i)\!-\!e^{-p_i} d_i(x_i)\right]\\ & +& x_1\mu \left(e^{p_2 - p_1} \!-\! 1\right)+ x_2\mu\alpha \left(e^{p_1 - p_2}\!-\! 1\right)\!, \nonumber
\end{eqnarray}
where $p_i=\partial_{x_i} S$ are the conjugate momenta. Hamiltonian~(\ref{HamiltonianDdim}) yields a set of four Hamilton equations, which can be solved numerically for any set of parameters, yielding the MTE~\cite{dykman2008disease,schwartz2009predicting}. Analytical progress can be made in two limiting cases: slow and fast migration.

\begin{figure}[t!]
	\includegraphics[width=0.8\linewidth]{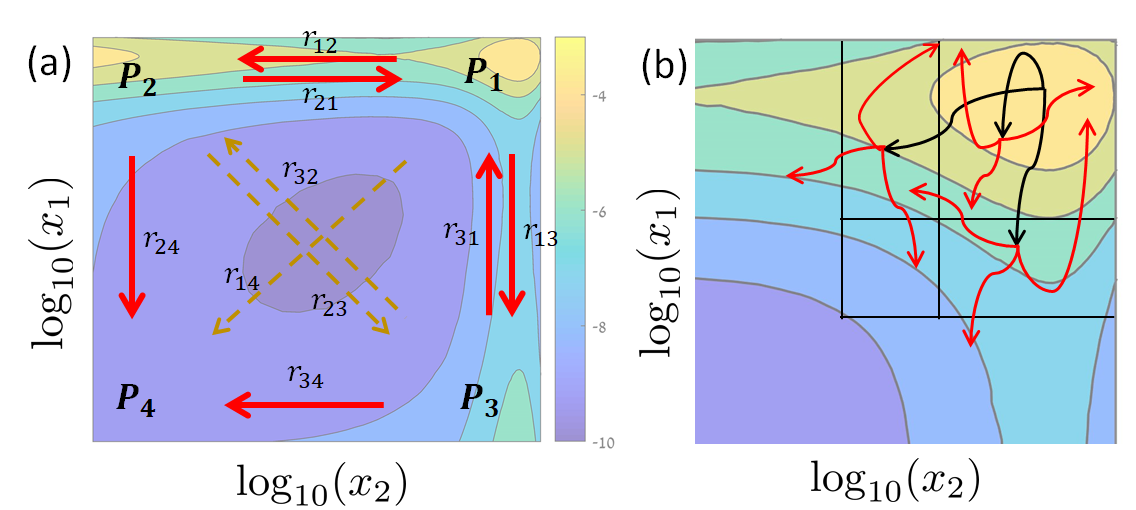}
\vspace{-2mm}
	\caption{Phase space and WE simulations. (a) Transitions between basins of attraction of FPs 1-4, see text, drawn over the quasistationary distribution obtained by the WE simulation~\cite{SM}. (b) Illustration depicting the two steps we repeat in a WE simulation, propagation and re-sampling, see~\cite{SM} for details. Parameters in both panels are $N_1 = N_2 = 100, d_1 = 0.5, d_2 = 0.4, \mu = 10^{-3}$.}
	\label{fig2}
\end{figure}

\textit{The case of slow migration.} For slow migration, $\mu\ll 1$, in general there are four stable FPs at the deterministic level. However, when accounting for demographic noise, these FPs become metastable states, which means that the system can stochastically switch between any pair of them. Importantly, the presence of multiple metastable states gives rise to multiple extinction routes. To find the MTE of the metapopulation, we apply a similar method to Ref.~\cite{gottesman2012multiple}, and define
$
\mathcal{P}_{1}\equiv\mathcal{P}(\{x_{1, +}, x_{2, +}\}) ,
\mathcal{P}_{2}\equiv\mathcal{P}(\{x_{1, +}, x_{2, 0}\}),
\mathcal{P}_{3}\equiv\mathcal{P}(\{x_{1, 0}, x_{2, +}\}),
\mathcal{P}_{4}\equiv\mathcal{P}(\{x_{1, 0}, x_{2, 0}\}) ,
$
as the probabilities to be at the basins of attraction of the FPs where both patches are colonized (FP1), patch 1 is colonized and patch 2 is extinct (FP2), patch 2 is colonized and patch 1 is exinct (FP3), and both patches are extinct (FP4); see Fig.~\ref{fig2}(a) for an illustration. Assuming the transition rates $r_{ij}$ between $\mathcal{P}_i$ and $\mathcal{P}_j$ are known (to be calculated below), the probabilities $\mathcal{P}_i$ ($i=1,2,3,4$) satisfy
\begin{eqnarray} \label{P1_4}
\dot{\mathcal{P}}_{i}\left(t\right)=  \sum_{j\neq i}r_{ji}\mathcal{P}_{j}(t)-r_{ij}\mathcal{P}_{i}(t),
\end{eqnarray}
where the MTE is given by $\tau=\int_{0}^{\infty}t\mathcal{\dot{P}}_4 dt$~\cite{gottesman2012multiple,r14}.
These equations can be solved for any $r_{ij}$~\cite{SM, horn2012matrix}. Yet, these transition rates, in general, exponentially differ from each other. Using this fact, the solution to $\tau$ simplifies to be:
\begin{eqnarray} \label{MTEgeneral}
&&\tau  \simeq  \text{min}\left\{\text{max}\left[r_{12}^{-1},r_{24}^{-1}, r_ {21}/(r_ {12} r_ {24})\right],\right.\nonumber\\ &&\quad\quad\quad\quad\left.\text{max}\left[r_{13}^{-1},r_{34}^{-1}, r_ {31}/(r_ {13} r_ {34}) \right]\right\},
\end{eqnarray}
where we have assumed that switches, which involve synchronous transitions of \textit{both} patches, occur at an exponentially slower rate than other transitions; \textit{i.e.} rates $r_{14}, r_{23}, r_{32}$, are negligible compared to all other rates~\cite{SM}. The outer minimum in~(\ref{MTEgeneral}) chooses the extinction route with the overall minimal cost, while the inner maximum determines the cost of the chosen trajectory.

\begin{figure}[t!]
\vspace{-3mm}
	\includegraphics[width=0.93\linewidth]{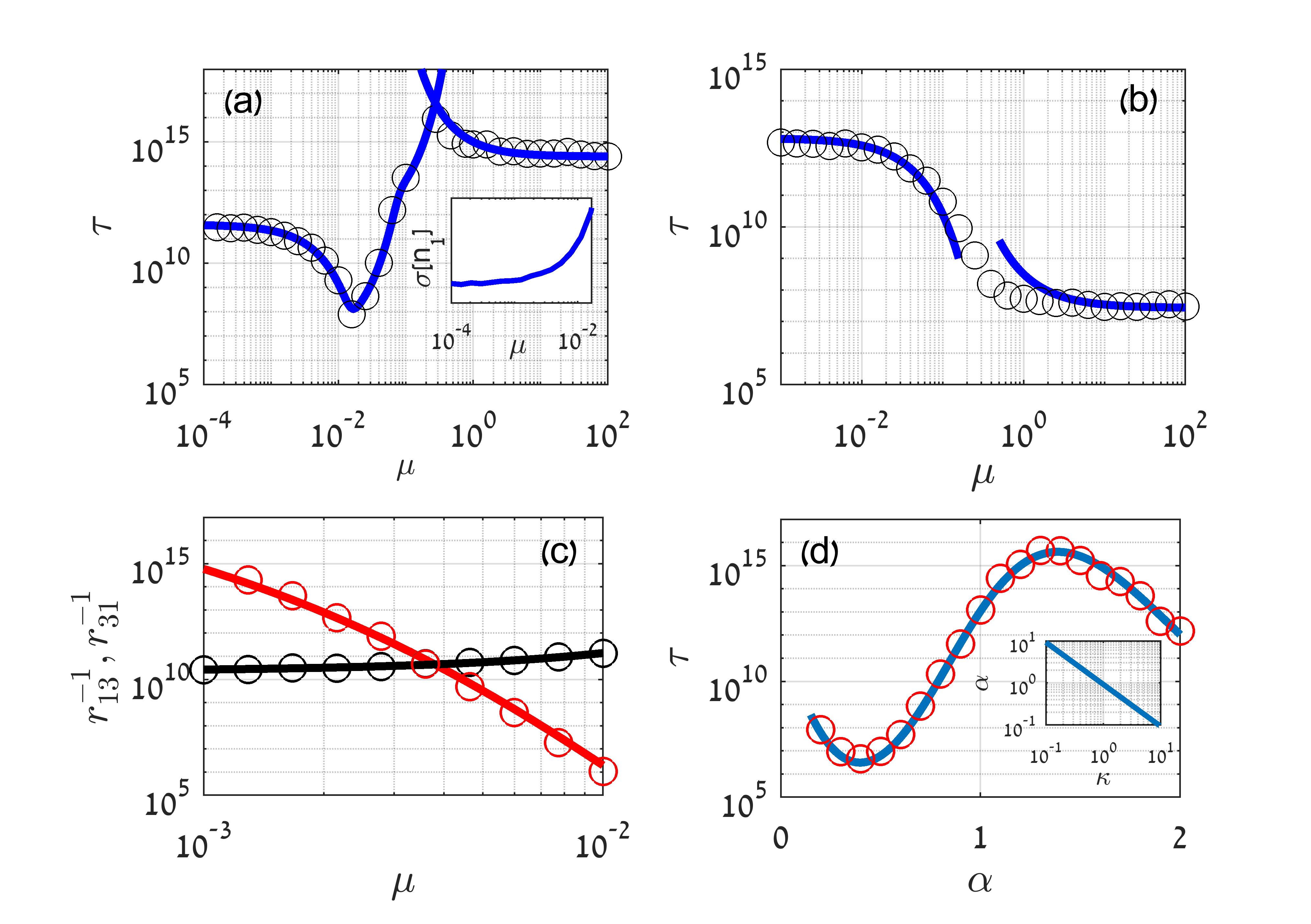}
\vspace{-5mm}
	\caption{ MTE as a function of $\mu$: theory (solid line) given by Eq. (\ref{MTEgeneral}) with Eqs. (\ref{Action_rij}) for slow migration, and Eq.~(\ref{ActionHighMigration2dim}) for fast migration (including correction in $1/\mu$), compared to WE simulations (circles). In (a)
$N_1 = 2300$, $N_2 = 2000$, $\delta_1 = 0.25$, $\delta_2 = 0.21$, $\alpha = 1$; in (b) $N_1 = 400$, $N_2 = 100$, $\delta_1 = 0.45$, $\delta_2 = 0.7$, $\alpha = 1$.  Inset of (a) shows the sharp increase in the standard deviation of patch 1 around its stable FP (normalized by the mean size of both patches) when approaching $\mu_{crit}\simeq 1.5\cdot 10^{-2}$.  In (c) we compare the extinction (black line) and colonization (red line) rates, for $N_1 = 250$, $N_2 = 125$, $\delta_1 = 0.62$, $\delta_2 = 0.6$, $\alpha = 1$. In (d) we show $\tau$ versus $\alpha$, while the inset shows the values of $\alpha$ and $\kappa$ that maximize $\tau$; here, $N_1 = 300$, $N_2 = 200$, $\delta_1 = 0.56$, $\delta_2 = 0.34$, $\mu = 100$.}
	\label{fig3}
\end{figure}

We now compute the rates $r_{ij}$ in the limit of $\mu\ll 1$, while taking into account the fact that in this limit, extinction occurs in a \textit{serial} manner with an overwhelming probability~\cite{SM}. Without loss of generality let us consider the extinction of patch 1 while patch 2 remains colonized. That is, we assume the population of patch 2 fluctuates about its stable (colonized) FP with fluctuations of ${\cal O}(N^{-1/2},\mu)$~\cite{SM}. To this end, we substitute $x_2 = \bar{x}_{2,+}[1 + {\cal O}(\mu)+{\cal O}(N^{-1/2})]$ and $p_2 = {\cal O}(\mu)+{\cal O}(N^{-1/2})$, into Hamiltonian (\ref{HamiltonianDdim}), and keep terms up to first order in $\mu \ll 1$ and zeroth order in $N\gg 1$. After some algebra, this yields an effective Hamiltonian that accounts for the transition FP1$\to$FP3, \textit{i.e.}, the extinction of patch $1$, while experiencing patch 2 as constant external flux at its colonized state [see Fig. \ref{fig2}(a)]:
\begin{eqnarray} \label{HamiltonianLowMigration}
H_{\text{eff}}^{\text{slow}}(x_1, p_1) &&=  \left(e^{p_1}-1\right)\left(b_1(x_1)-e^{-p_1} d_1(x_1) \right)  \\ && +   \mu \left(e^{-p_1}-1\right) x_1+  \alpha\mu \kappa(1+\delta_2)\left(e^{p_1}-1\right). \nonumber
\end{eqnarray}
From this Hamiltonian we can compute the optimal path along the transition FP1$\to$FP3 and the corresponding action $S_{13}$. The actions along the transition paths FP1$\to$ FP2, FP2$\to$ FP4, and FP3$\to$ FP4, can be computed in a similar manner. The results are~\cite{SM}:
\begin{eqnarray}\label{Action_rij}
\hspace{-1mm}&&S_{24}\! =\! S_0(\delta_1)(1\!+\!\mu/2)\! -\! \mu \delta_1,\;\;S_{13}\! =\! S_{24}\!+\!\mu\alpha\kappa\delta_1(1\!+\!\delta_2), \\
\hspace{-1mm}&&S_{34}\! =\! \kappa S_0(\delta_2)(1\! +\! \mu \alpha/2)\!  -\! \mu\alpha \kappa \delta_2,\;\;S_{12}\!=\!S_{34}\!+\!\mu\delta_2(1\!+\!\delta_1) \nonumber,
\end{eqnarray}
where $S_0(\delta_i) =  2 \left[\delta_i - (1 - \delta_i^2)^{1/2} \arcsin(\delta_i) \right]$ is the action of isolated patch $i$.
Given these actions, the transition rates are given by $r_{ij} = e^{-N S_{ij}}$ for $ij = \{12, 13, 34, 24\}$. In addition, the \textit{colonization} rates $r_{21}$ and $r_{31}$, can also be found using Hamiltonian~(\ref{HamiltonianLowMigration}), see Fig.~\ref{fig3}(c)~\cite{SM}.

Having found all possible transition rates that contribute to the MTE, $\tau$, see Eq.~(\ref{MTEgeneral}), we are now in the position to study the latter as a function of the migration rate. In Fig.~\ref{fig3}(a-b) we compare, for two different parameter sets, our analytical result for $\tau$ versus $\mu$ [Eq.~(\ref{MTEgeneral})], with highly-efficient WE
simulations~\cite{huber1996weighted, zhang2010weighted, zuckerman2017weighted}, see Figs.~\ref{fig2}(b),~S7,~S8, and~\cite{SM} for details.
In Fig.~\ref{fig3}(a-b), for sufficiently slow migration, one observes a decrease in $\tau$ as $\mu$ increases, see also Figs.~S2 and~S3. That is, as $\mu$ is increased from zero, the metapopulation's extinction risk increases, compared to the isolated case.

To understand why this occurs, let us analyze the competing terms in Eq.~(\ref{MTEgeneral}). For $\mu\to 0$, the colonization rates $r_{21}, r_{31}$ vanish, and thus, at sufficiently slow migration, these rates can be neglected. Moreover, while at $\mu=0$, we have $r_{13}=r_{24}$ and $r_{12}=r_{34}$, see Eqs.~(\ref{Action_rij}), as $\mu$ is increased, $r_{24}$ and $r_{34}$ respectively increase at a faster rate than $r_{13}$ and $r_{12}$ (which do not necessarily increase at all). Thus, as $\mu$ increases, the minimum in Eq.~(\ref{MTEgeneral}) necessarily chooses the terms $r_{24}^{-1}$ over $r_{13}^{-1}$ and $r_{34}^{-1}$ over $r_{12}^{-1}$. As a result, the MTE is determined by the maximum of $r_{24}^{-1}$ and $r_{34}^{-1}$, both of which decrease as $\mu$ is increased.

Why do $r_{24}^{-1}$ and $r_{34}^{-1}$ decrease? Along the transitions FP2$\to$FP4 and FP3$\to$FP4 there is one patch that is colonized and another, close to extinction. As $\mu$ grows, the colonized patch sends individuals to the patch close to extinction, while the back flux is negligible. Thus, whereas the flux from the colonized patch increases its extinction risk (due to loss of individuals), it cannot rescue the other patch from extinction, as its population size is below the colonization threshold. Hence, weak migration increases the metapopulation's global extinction risk compared to the isolated case, which is a direct consequence of the Allee effect and the existence of a colonization threshold; for local logistic dynamics the opposite is observed~\cite{khasin2012minimizing}.

The decrease of $\tau(\mu)$ at small $\mu$ can give rise to another fascinating phenomenon: the existence of a global minimum of $\tau(\mu)$ at a critical (and finite) migration rate $\mu_{crit}$, that maximizes the extinction risk of the metapopulation~\footnote{The general conditions for the existence of such a  minimum at a finite $\mu$ is given in \cite{SM}, see also Fig.~S5.}. In Fig.~\ref{fig3}(a) we observe such a global minimum, while in Fig.~\ref{fig3}(b), the minimum is obtained only at $\mu\to\infty$. Empirically, this minimum is accompanied by a sharp increase in the population's variance of the colonized patch (in FP2/FP3) as $\mu$ approaches $\mu_{crit}$, see inset of Fig.~\ref{fig3}(a). This increase can serve as an early warning signal for stability deterioration of the population~\cite{scheffer2009early}.

Notably, our analysis invalidates a long standing claim, that the rescue of local patches necessarily increases the metapopulation's stability~\cite{hanski1999metapopulation, eriksson2014emergence}, when the local dynamics include the Allee effect. Indeed, for small $\mu$, as $\mu$ increases a single patch can experience a decrease in its local extinction risk (``rescue effect") even though the global extinction risk increases, see Fig.~\ref{fig3}(c) and Fig.~S4.

\textit{The case of fast migration.} In the limit of fast migration, $\mu\!\gg\! 1$, only FPs 1 and 4 remain, see Figs.~\ref{fig1} and~\ref{fig2}. Therefore, extinction can only occur via a transition between a metastable state where both patches are colonized, and the extinction state, and the MTE is given by $r_{14}^{-1}$. To find the action, we exploit the fact that in this limit, the total population size in both patches is slowly varying compared to the local population on each patch. Using a transformation of variables
$Q=  x_1 + x_2$, $q=x_2$, $P=(p_1 + p_2)/2$ and $p=p_2$, which is canonical up to ${\cal O}(\mu^{-1})$~\cite{SM}, and performing adiabatic elimination of the fast variables $q$ and $p$~\cite{assaf2008noise}, we arrive at an effective Hamiltonian for the slow variables $Q$ and $P$~\cite{khasin2012fast}
\begin{eqnarray} \label{HamiltonianHighMigrationDdim}
H_{\text{eff}}^{\text{fast}} =  \left(e^{P}-1\right) && \left[\tilde{b}(Q) -e^{-P} \tilde{d}(Q)\right],
\end{eqnarray}
with $\tilde{b}(x) = 2 x^2/[\tilde{\kappa}(1+1/\alpha)(1 - \tilde{\delta}^2)]$ and $\tilde{d}(x) =  x + x^3/[\tilde{\kappa}^2(1+1/\alpha)^2(1 - \tilde{\delta}^2)]$.
Integrating along the optimal path, one arrives at the total action in this case~\cite{SM}
\begin{equation} \label{ActionHighMigration2dim}
S^{\text{fast}} = (1+1/\alpha) \tilde{\kappa } S_0(\tilde{\delta}).
\end{equation}
Note that we have also computed the subleading ${\cal O}(\mu^{-1})$ correction to $S^{\text{fast}}$ in the limit of fast migration, see~\cite{SM} for a detailed discussion and also Fig.~S6.
Also note that, upon replacing the threshold and carrying capacity parameters by their effective counterparts, $S^{\text{fast}}$ coincides with the one-patch result, up to a factor of $(1+1/\alpha)$, which corresponds to the combined (deterministic) contribution of the two patches. Importantly, this leading-order result in $\mu\gg 1$, provides an indication whether fast migration is beneficial over isolation for the entire metapopulation. That is,
if $S^{\text{fast}} > \max\left\{S_0(\delta_1), \kappa S_0(\delta_2)\right\}$~\cite{Scond}, fast migration has a positive effect on the population's viability. In this case, in addition to $\mu_{crit}$ for which the MTE is minimized, there exists an optimal migration rate, $\mu_{opt}$, which globally maximizes the MTE, see Figs.~\ref{fig3}(a) and~S2. In contrast, in Figs.~\ref{fig3}(b) and~S3, this condition does not hold, and $\tau$ decreases monotonically for the entire range of $\mu$.

Finally, it can be shown that when $\delta_1$ and $\delta_2$ are comparable, the fast-migration action (\ref{ActionHighMigration2dim}) is maximized when $\alpha \simeq 1/\kappa$, see Fig.~\ref{fig3}(d). This is another main result of this work: at fast migration, when the typical flux between patches is approximately equal, the extinction risk is minimized. When this condition is met, typically an equal number of individuals pass across patches per unit time, which corresponds to an optimal synchronization of patches. In contrast, when this condition is not met, and $\alpha\kappa$ significantly deviates from $1$, the synchronization breaks down and one patch becomes significantly less stable than the other, resulting in a much lower MTE than the synchronized case. In extreme cases, this loss of synchrony may lead to a ``source-sink" dynamics where the patch with the large carrying capacity becomes a sink to the patch with the small carrying capacity~\cite{SM}.

The analysis above is not limited to our particular choice of the birth and death rates [Eq.~(\ref{rates})], and is generic for any model, locally exhibiting the Allee effect~\cite{SM,mendez2019demographic}. Notably, our approach can be used to analyze multi-state gene regulatory networks, where each ``patch" corresponds to a distinct DNA state. Here the local Allee-like dynamics of proteins is supplemented by protein influx such that instead of extinction, the system switches between different phenotypic states~\cite{choi2008stochastic,assaf2011determining,earnest2013dna}. Our method allows rigorous treatment of such models in the important limits of fast and slow binding/unbinding of a repressor/promoter to the DNA states, compared to protein synthesis/degradation~\cite{hornos2005self,morelli2009dna}. Moreover, our approach may provide insight into the dynamics of a bacterial population under antibiotic stress, where it has been observed that demographic fluctuations can be reduced by migration between the two ``patches", corresponding to the persister and non-persister phenotypic states~\cite{pearl2008nongenetic}.

We thank Michael Khasin and Yonatan Friedman for useful discussions. We acknowledge support from the Israel Science Foundation grant No. 300/14 and the United States-Israel Binational Science Foundation grant No. 2016-655.

\section*{\Large{Supplemental Material}}

\renewcommand{\thefigure}{S\arabic{figure}}
\renewcommand{\theequation}{S\arabic{equation}}
\setcounter{figure}{0}   
\setcounter{equation}{0}

\section{Deterministic dynamics}
In this section we derive the deterministic fixed points (FPs) for slow and fast migration. We begin with the rate equations for $M$ patches [Eqs.~\eqref{RateEqMultiPatch_x} in the main text]:
\begin{equation} \label{RateEqMultiPatch_x_Mpatches}
\dot{x_i}=  b_i(x_i) - d_i(x_i) + \mu \sum_{j \neq i}^M (\alpha_{ji} x_j -  \alpha_{ij} x_i),
\end{equation}
where $x_i$ is the population density at patch $i$, and we have denoted by $\mu_{ij} = \alpha_{ij}\mu$  the migration rate between patch $i$ and $j$, while $\alpha_{ij} = \mathcal{O}(1)$. For two patches, $M=2$, we denote for simplicity $\alpha_{12} = 1$ and $\alpha_{21} = \alpha$, and the rate equations can be explicitly written as:
\begin{eqnarray} \label{RateEqMultiPatch_xSM}
&&\dot{x}_1=  b_1(x_1) - d_1(x_1) + \alpha \mu x_2 -  \mu x_1 \\
&&\dot{x}_2=  b_2(x_2) - d_2(x_2) + \mu x_1 -  \alpha \mu x_2 , \nonumber
\end{eqnarray}
where $b_i(x_i)$ and $d_i(x_i)$ are given by
\begin{equation} \label{birth_death}
b_i(x_i) \equiv  \frac{2 x_i^2}{\kappa_i (1-\delta_i^2)}  \;\; , \;\; d_i(x_i) = x_i + \frac{x_i^3}{\kappa_i^2 (1-\delta_i^2)}.
\end{equation}
Here $0<\delta_i < 1$ determines the distance between the local colonized and colonization threshold states,  $N_i\gg 1$ is the local carrying capacity, and $\kappa_i=N_i/N_1$ such that $\kappa_1=1$ and $\kappa_2 \equiv \kappa={\cal O}(1)$, see main text.
In Fig.~\ref{fig1sm} (see also Fig.~1 in the main text) we show examples of numerical solutions of Eqs.~\eqref{RateEqMultiPatch_xSM}, see below.

\subsection{The case of slow migration}
\begin{figure}[t]
	\includegraphics[width=1\linewidth]{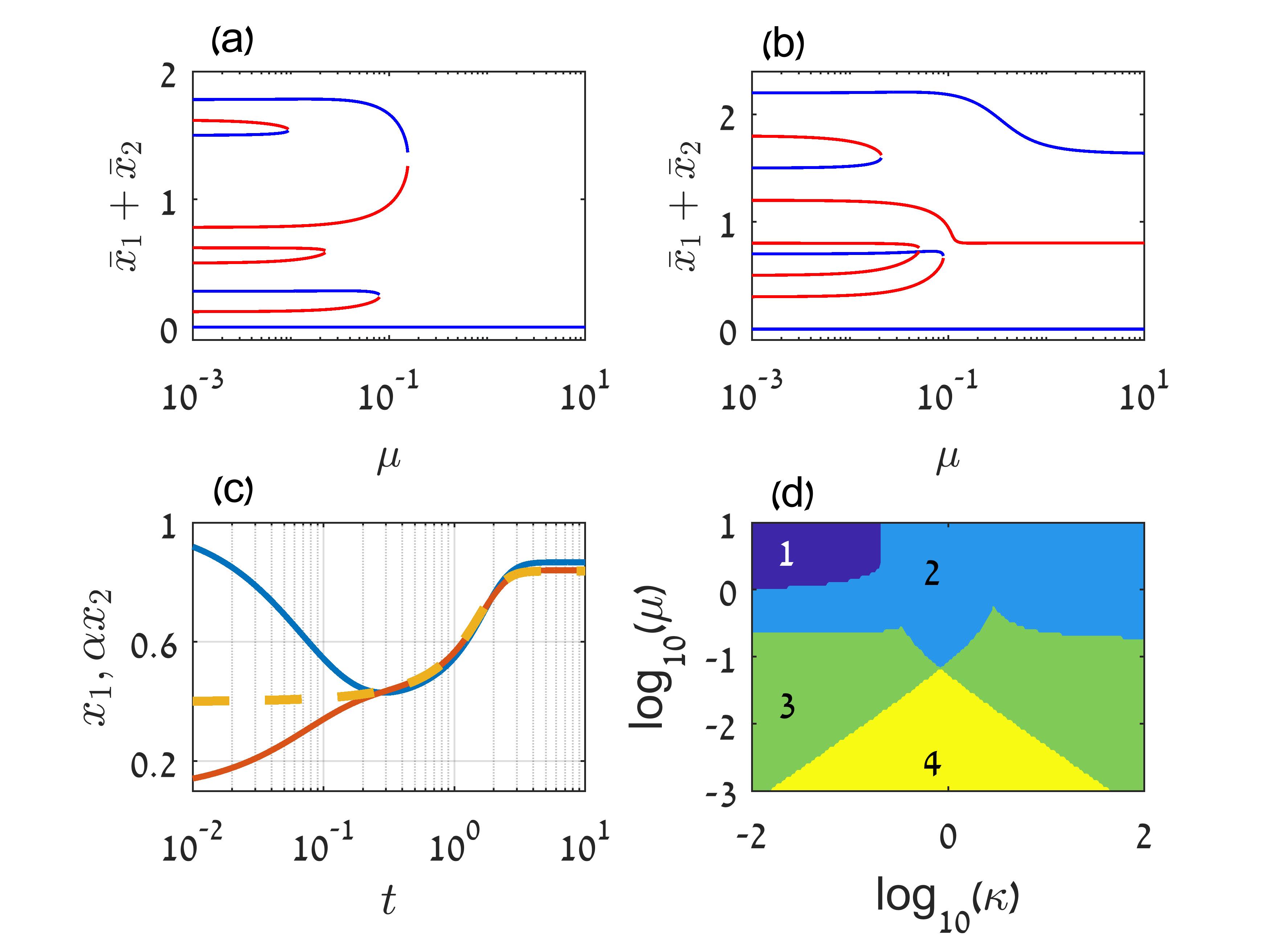}
	\caption{ (a) and (b): Bifurcation diagrams of Eqs.~(\ref{RateEqMultiPatch_xSM}), as a function of $\mu$. In (a) $\kappa = 0.2, \delta_1 = 0.5, \delta_2 = 0.4 , \alpha = 1$, while in (b) $\kappa = 0.5, \delta_1 = 0.5, \delta_2 = 0.4 , \alpha = 1$. Here, the blue and red lines correspond to stable and unstable FPs, respectively. (c): $x_1$ and $ \alpha x_2 $  as a function of time by numerically solving Eqs.~(\ref{RateEqMultiPatch_xSM}) (blue and red lines), compared to numerical solution of Eq.~(\ref{RateEqMultiPatch_High}) for $\xi$ (dashed line). Parameters are $\kappa = 1, \delta_1 = 0.5, \delta_2 = 0.6, \mu = 10, \alpha = 0.5$. (d): Number of stable FPs as function of both $\kappa$ and $\mu$, for $\delta_1 = 0.71, \delta_2 = 0.4$, $\alpha = 1$. }
	\label{fig1sm}
\end{figure}

At zero migration each patch has three FPs. These are found by putting $\dot{x}_1=\dot{x}_2=0$ in Eqs.~(\ref{RateEqMultiPatch_xSM}) with $\mu = 0$:
\begin{eqnarray} \label{FP_isolated}
&&\bar{x}^{(0)}_{i, 0} = 0\;, \;\; \bar{x}^{(0)}_{i, \pm} = \kappa_i (1 \pm \delta_i),
\end{eqnarray}
where $\bar{x}^{(0)}_{i, 0}$ and $\bar{x}^{(0)}_{i, +}$ are stable and $\bar{x}^{(0)}_{i, -}$ is unstable. As a result, there are nine ($3^2$) FPs for zero migration. To find the FPs for $\mu\ll 1$, we look for the solution as $\bar{\bm{x}}= (\bar{x}_1, \bar{x}_2) = (\bar{x}^{(0)}_{1, s_1}+\mu \bar{\eta}_{1, \bm{s}}, \bar{x}^{(0)}_{2, s_2}+\mu \bar{\eta}_{2, \bm{s}})$ for $\bm{s} = (s_1, s_2)$ with $s_i = \{0, +, - \}$ representing the possible states, where $\bar{x}^{(0)}_{i, s_i}$ are given by Eq.~\eqref{FP_isolated} and $\bar{\eta}_{i, \bm{s}}$ are yet unknown. Substituting this solution into Eqs.~\eqref{RateEqMultiPatch_xSM}, putting $\dot{x}_1=\dot{x}_2=0$ and keeping terms up to ${\cal O}(\mu)$, we find $\bar{\eta}_{i, \bm{s}}$, which yields the FPs up to ${\cal O}(\mu)$:
\begin{eqnarray} \label{x_mf_lowMigration}
\bar{\bm{x}}_{\bm{s}} =&& \left[\bar{x}_{1,s_1}^{(0)}+\mu t^{relax}_{1, s_1}\left(\alpha\bar{x}^{(0)}_{2, s_2} - \bar{x}^{(0)}_{1,s_1} \right) ,  \right. \\ &&\left. \bar{x}_{2,s_2}^{(0)}+\mu t^{relax}_{2, s_2} \left(\bar{x}^{(0)}_{1, s_1}-\alpha \bar{x}^{(0)}_{2,s_2}\right) \right]. \nonumber
\end{eqnarray}
Here we have defined $t_{i,s_i}^{relax} = [b_i'(\bar{x}_{i, s_i}^{(0)}) - d_i'(\bar{x}_{i, s_i}^{(0)})]^{-1}$,  as the relaxation time to the FP $\bar{x}_{i, s_i}^{(0)}$, at the level of the isolated patch.
This result [Eq.~\eqref{x_mf_lowMigration}] can be intuitively understood as follows: in the leading order the average population density in each patch is determined by the outgoing flux from itself and incoming flux from the second patch. The magnitude of the correction depends on the migration rate as well as the relaxation time to the relevant FP.
Note that since $\bm{s} = (s_1, s_2)$ can receive nine different values, Eq.~\eqref{x_mf_lowMigration} represents $3^2=9$ FPs. It can be shown via linear stability analysis that four of these FPs are stable while five of them are unstable. The former correspond to scenarios where either both patches are colonized, one patch is colonized and the other is close to extinction, and both patches are extinct. This can be observed in Fig.~\ref{fig1sm}(a-b) and also in Fig.~1 in the main text. One can see that as $\mu$ is increased, the number of FPs decreases, see also Fig.~\ref{fig1sm}(d) and below.

For clarity let us give two  examples of Eq.~(\ref{x_mf_lowMigration}). In the case where  patch $1$ is colonized and $2$ is close to extinction, the stable FP is given by:
\begin{eqnarray} \label{patch1colonized}
\bar{\bm{x}}_{+,0}&&=[(\bar{\bm{x}}_{+,0})_1,(\bar{\bm{x}}_{+,0})_2] \\ &&=  [1 + \delta_1 -\mu (1-\delta_1^2)/(2\delta_1)\;,\; \mu (1 + \delta_1)], \nonumber
\end{eqnarray}
while in the opposite case where patch $2$ is colonized and $1$ is close to extinction, the stable FP is given by:
\begin{eqnarray} \label{patch2colonized}
\bar{\bm{x}}_{0,+}&&=[(\bar{\bm{x}}_{0,+})_1,(\bar{\bm{x}}_{0,+})_2] \\ &&= [ \mu \alpha \kappa (1 + \delta_2) \;,\; \kappa(1+\delta_2)  - \mu\alpha\kappa(1 -\delta_2^2)/(2\delta_2)] .\nonumber
\end{eqnarray}
Here we have used the fact that $t_{i,+}^{relax} = (1-\delta_i)/(2\delta_i)$ and $t_{i,0}^{relax} =1 $, while the notation $(\bar{\bm{x}}_{\bm{s}})_i$ indicates element $i=1,2$ of vector $\bar{\bm{x}}_{\bm{s}}$, see Eq.~\eqref{x_mf_lowMigration}.
In Eqs.~\eqref{patch1colonized} and ~\eqref{patch2colonized} it is evident that the correction to the colonized patch is necessarily negative, as deterministically, the incoming flux from the patch close to extinction contributes only ${\cal O}(\mu^2)$ terms.

\subsection{The case of fast migration}
As $\mu$ increases the number of FPs decreases via multiple bifurcations, see Fig.~\ref{fig1sm}(a-b) and below. For fast migration, $\mu\gg 1$, there are at most three FPs, two of which are stable.
To see this, we put $\dot{x}_1 = \dot{x}_2 = 0$ in Eqs.~(\ref{RateEqMultiPatch_xSM}), and in the leading order, we neglect all terms that do not depend on $\mu$. This results in $\mu (x_1-\alpha x_2)=0$, suggesting that patches are synchronized via $x_2 = x_1/\alpha $. To analyze the dynamics in the fast migration limit, we sum Eqs.~(\ref{RateEqMultiPatch_xSM}) and substitute $x_1 = \xi, x_2 = \xi/\alpha$. This yields an effective rate equation for $\xi$
\begin{equation} \label{RateEqMultiPatch_High}
\dot{\xi}=-\xi +  \frac{2 \xi^2}{\tilde{\kappa} (1-\tilde{\delta}^2)}  -\frac{ \xi^3}{\tilde{\kappa}^2 (1-\tilde{\delta}^2)},
\end{equation}
where $\tilde{\kappa}$ and $\tilde{\delta}$ are given by
\begin{eqnarray} \label{EffectiveParameters}
&&\tilde{\kappa} = \frac{\alpha  \kappa  \left[\alpha^2\kappa  (1-\delta _2^2) +1-\delta _1^2\right]}{\alpha^3 \kappa^2 (1-\delta _2^2) + 1-\delta _1^2}\nonumber\\
&& \tilde{\delta} = \left[1 - \frac{\alpha  (\alpha +1) \kappa (1-\delta _1^2 ) \left(1-\delta _2^2\right)}{\tilde{\kappa}  \left[\alpha ^2 \kappa(1-\delta _2^2) + 1 - \delta_1^2\right]}\right]^{1/2},
\end{eqnarray}
and $\tilde{\kappa} >0$ by definition, see main text.
In Fig.~\ref{fig1sm}(c) we demonstrate the synchrony between $x_1$ and $\alpha x_2$ by numerically solving Eqs.~(\ref{RateEqMultiPatch_xSM}) and comparing it with a numerical solution to Eq.~(\ref{RateEqMultiPatch_High}). As can be observed in this panel, convergence of the two patches is achieved at time $t\gtrsim\OO(\mu^{-1})$, making Eq.~\eqref{RateEqMultiPatch_High} valid at times $t \gg \OO(\mu^{-1})$.

A sufficient condition for bistability in Eq.~\eqref{RateEqMultiPatch_High} is that $0<\tilde{\delta}<1$ be real, which gives rise to three FPs:
\begin{equation} \label{mf_high_migration}
\xi_0 = 0\;,\;  \xi_{\pm} = \tilde{\kappa}(1\pm \tilde{\delta}).
\end{equation}
These FPs coincide with those in Eqs.~(\ref{FP_isolated}), upon replacing $\delta$ and $\kappa$ by their effective counterparts, see Eqs.~\eqref{EffectiveParameters}.

A simple example demonstrating this case is when $\delta_1=\delta_2=\delta$ and $\kappa = \alpha^{-1}$. This yields $\tilde{\kappa}= 1$ and $\tilde{\delta}=\delta$, and thus, the FPs in Eqs.~\eqref{mf_high_migration} become $\xi_\pm= (1 \pm \delta)$. This is not a trivial result as it predicts that when $\alpha$ and $\kappa$ counter each other such that the typical flux between patches is approximately equal, the resulting dynamics mimic those of the original patches. In this case, the system is said to be well mixed [12].

However, if $\tilde{\delta}$ is complex, $\xi=0$ is the only stable FP, and $\xi$ decays to $0$ deterministically. This scenario can occur, \textit{e.g.}, when the carrying capacities are very different, $\kappa \ll 1$, where $\alpha = \OO(1)$. In this case, we find in leading order of $\kappa$, $\tilde{\kappa} \simeq \kappa \alpha$ and $\tilde{\delta} \simeq [\delta_2^2 - \alpha (1-\delta_2^2)]^{1/2}\equiv \Delta_2$. Since these only depend on the parameters of patch 2, which has a much smaller carrying capacity, this entails that the smaller patch dictates the deterministic size of the population. Thus, the system will maintain a colonized FP only as long as $\Delta_2$ is real, which yields
$\delta_2 \geq [\alpha/(1+\alpha)]^{1/2}$.  The same behavior is observed for $\kappa \gg 1$, where in this case the stability is determined by the parameters of patch 1; here, the existence of a colonized state requires $\delta_1\geq [1/(1+\alpha)]^{1/2}$.

The fact that the system can be either bistable or monostable at large $\mu$ is demonstrated in Fig.~\ref{fig1sm} (see also Fig.~1 in the main text). In Fig.~\ref{fig1sm}(d) [see also Fig.~1(d) in the main text] we map the number of stable FPs as a function of both $\mu$ and $\kappa$; here, as $\mu$ is increased, the number of stable FPs tends either to 1 or 2 depending on the value of $\kappa$, and the other parameters. For $\kappa \ll 1$ and $\delta_2  < [\alpha/(1+\alpha)]^{1/2}$, see above, we get deterministic extinction, and similarly for $\kappa \gg 1$ and $\delta_1 < [1/(1+\alpha)]^{1/2}$; otherwise, the dynamics is bistable.

In Fig.~\ref{fig1sm}(a-b) [see also Fig.~1(c) in the main text], by numerically solving Eqs.~(\ref{RateEqMultiPatch_xSM}) for the entire range of $\mu$, we demonstrate the multiple bifurcations occurring as $\mu$ increases. Starting from nine FPs at small $\mu$, as $\mu$ is increased, the system ends up with either one or three FPs, corresponding to deterministic extinction, and a bistable system with a long-lived colonized state, respectively. As shown in Fig.~\ref{fig1sm}(a,b,d), the number of stable FPs at large $\mu$ strongly depends on $\kappa$.

Finally, we can compute the subleading-order corrections in $\zeta=\mu^{-1}$ to the FPs in the limit of $\mu\gg 1$. That is, we can find the $\OO(\zeta)$ corrections to the FPs given by Eq~\eqref{mf_high_migration}. While we do not give the explicit expressions here, these will be used to compute the subleading-order corrections to the mean time to extinction (MTE) at fast migration, see below.

\section{Stochastic dynamics}
\subsection{Slow migration - multiple extinction routes}
In this subsection we derive the MTE [Eq.~(\ref{MTEgeneral}) in the main text].
Our staring point are Eqs.~(\ref{P1_4}) in the main text:
\begin{eqnarray}
&&\dot{\mathcal{P}}_{1}\left(t\right)=  r_{21}\mathcal{P}_{2}\left(t\right)+r_{31}\mathcal{P}_{3}\left(t\right)-\left(r_{12}+r_{13}+r_{14}\right)\mathcal{P}_{1}\left(t\right)    \nonumber\\
&&\dot{\mathcal{P}}_{2}\left(t\right)=  r_{12}\mathcal{P}_{1}\left(t\right)+r_{32}\mathcal{P}_{3}\left(t\right)  -\left(r_{24}+r_{21}+r_{23}\right)\mathcal{P}_{2}\left(t\right)   \nonumber \\
&&\dot{\mathcal{P}}_{3}\left(t\right)= r_{13}\mathcal{P}_{1}\left(t\right)+r_{23}\mathcal{P}_{2}\left(t\right) -\left(r_{34}+r_{31}+r_{32}\right)\mathcal{P}_{3}\left(t\right)  \nonumber \\
&&\dot{\mathcal{P}}_{4}\left(t\right)=   r_{14}\mathcal{P}_{1}\left(t\right)+r_{24}\mathcal{P}_{2}\left(t\right)+r_{34}\mathcal{P}_{3}\left(t\right), \label{P1_4SM}
\end{eqnarray}
where $\mathcal{P}_i$ is the probability to be at the basins of attraction of FP $i$, see Fig.~2 in the main text, and $r_{ij}$ is the transition rate between $\mathcal{P}_i$ and $\mathcal{P}_j$.
Using the last of Eqs.~(\ref{P1_4SM}), the MTE given by $\tau =\int_{0}^{\infty}t\dot{\mathcal{P}}_{4}dt$, reads
\begin{eqnarray} \label{MTEdefinition2}
\tau =\int_{0}^{\infty}t\left[r_{14}\mathcal{P}_{1}(t) +r_{24}\mathcal{P}_{2}(t)+r_{34}\mathcal{P}_{3}(t)\right]dt .
\end{eqnarray}
Now, since $\mathcal{P}_4(t)$ does not appear explicitly in Eqs.~\eqref{P1_4SM}, we define $\bm{\mathcal{P}}(t) = [\mathcal{P}_1(t), \mathcal{P}_ 2(t), \mathcal{P}_3(t)]$ and rewrite the first three of Eqs.~\eqref{P1_4SM} in matrix form:
\begin{equation} \label{MultipleRoutesMatrixForm}
\dot{\bm{\mathcal{P}}}(t)= \matarrow{A} \bm{\mathcal{P}}(t) ,
\end{equation}
with
\begin{equation}
\matarrow{A} \equiv \left(\begin{array}{ccc}
-\sum_{i \neq 1}^4 r_{1i} & r_{21} & r_{31} \\
r_{12} & -\sum_{i \neq 2}^4 r_{2i} & 0 \\
r_{13} & 0 & -\sum_{i \neq 3}^4 r_{3i} \\
\end{array}\right).
\end{equation}
Note that in matrix $\matarrow{A}$ we have neglected the rates $r_{14}$, $r_{23}$, and $r_{32}$. Since at slow migration extinction occurs in a serial manner, it can be shown that in the semi-classical limit where the carrying capacities are large, these rates are negligible compared to other rates, as they satisfy $r_{14}\sim r_{12}r_{24}$, $ r_{23}\sim r_{21}r_{13}$, and $ r_{32}\sim r_{31}r_{12}$, see below.

In order to find the MTE we need to solve~\eqref{MultipleRoutesMatrixForm} with initial conditions $\bm{\mathcal{P}}(0) = (1,0,0)$, \textit{i.e.}, starting from the colonized state in both patches. Since $\matarrow{A}$ is not necessarily diagonalizable, the problem can be generally solved via the Schur decomposition [32], namely: $\matarrow{A} = \matarrow{U} \matarrow{T} \matarrow{U}^{-1}$, with $\matarrow{U}$ being a unitary matrix and $\matarrow{T}$ an upper triangular matrix, given by:
\begin{equation}
\matarrow{T} \equiv \left(\begin{array}{ccc}
\lambda_1 & T_{12} & T_{13} \\
0 & \lambda_2 & T_{23} \\
0 & 0 & \lambda_3 \\
\end{array}\right).
\end{equation}
Here, the diagonal elements $\lambda_i< 0$, are the eigenvalues of $\matarrow{A}$, and $T_{12},T_{13},T_{23}$ depend on the chosen (not unique) Schur decomposition. By decomposing $\bm{\mathcal{P}}(t) = \matarrow{U} \bm{q}(t)$ we rewrite Eq.~(\ref{MultipleRoutesMatrixForm}) as $\dot{\bm{q}}(t) = \matarrow{T} \bm{q}(t)$, with initial conditions $\bm{q}(0) = \matarrow{U}^{-1} \bm{\mathcal{P}}(0)$. Since $\matarrow{T}$ is upper triangular, the solution to $\bm{q}(t)= [q_1(t), q_2(t), q_3(t)]$ can be found iteratively by solving for $q_3(t)$, then $q_2(t)$, and then $q_1(t)$, as follows:
\begin{eqnarray}
q_3(t) &=& q_3(0) e^{\lambda_3 t},\\
q_2(t) &=&  q_2(0)e^{\lambda _2 t} +\int_0^t T_{23}   e^{-\lambda _2 (t'-t)} q_3(t') \, dt',\nonumber\\
q_1(t) &=&  q_1(0)e^{\lambda _1 t}\!+\!\int_0^t \! e^{-\lambda _1 (t'-t)} \left[T_{12} q_2(t')+T_{13} q_3(t')\right] \! dt',\nonumber
\end{eqnarray}
where these integrals can be computed in a straightforward manner. Finally, having found $\bm{q}(t)$, Eq.~\eqref{MTEdefinition2} is solved by
\begin{equation} \label{MTEexact}
\tau = \bm{r}_4^{\;\intercal} \matarrow{U} \bm{c}
\end{equation}
where $ \bm{r}_4 = (0, r_{24}, r_{34})$ and $\bm{c}= \int_0^\infty t \bm{q}(t) dt$ ($i = 1,2,3$). Note that while the Schur decomposition and the solution for $\bm{q}$ are not unique, the solutions for $\bm{\mathcal{P}}$ and $\tau$ [Eq.~\eqref{MTEexact}] are in fact unique.

While an explicit version of Eq.~\eqref{MTEexact} exists, it is highly cumbersome, as it involves solving a cubic equation and finding a specific schur decomposition.
Nonetheless, in one important case $\tau$ drastically simplifies.
In general, if the carrying capacities are large, the transition rates exponentially differ from each other. In particular, one of the transition rates from FP1 to FP2 or from FP1 to FP3, respectively given by $r_{12}$ and $r_{13}$ (see Fig.~2 in the main text), is negligible compared to the other.  Without loss of generality, let us assume that $r_{13}$ is negligible compared to $r_{12}$. In this case, the solution to Eq.~(\ref{MultipleRoutesMatrixForm}) drastically simplifies and reads
\begin{equation} \label{MTEapprox2}
\tau = \frac {r_ {12} + r_ {21} + r_ {24}} {r_ {12} r_ {24}}
\simeq \text{max}\left\{\frac {1}{r_ {24}},\frac{1}{r_ {12}},\frac {r_ {21}} {r_ {12} r_ {24}}\right\}.
\end{equation}
While this result only accounts for one route to extinction (assuming the other route is exponentially unlikely), it also includes a correction to this single route due to possible recolonization, which depends on the colonization rate $r_ {21}$. Demanding a-posteriori that $r_{13}$ be negligible compared to the individual rates comprising this route to extinction, and using Eq.~\eqref{Action_rij} in the main text, it can be shown that Eq.~\eqref{MTEapprox2} is valid as long as $r_{13} \ll r_{12} r_{24}/r_{21}$; otherwise we have neglected a term larger than the rate of extinction. Note that if alternatively $r_{13} \gg r_{12}$ then the equivalent of Eq.~\eqref{MTEapprox2} will hold as long as $r_{12} \ll r_{13} r_{34} / r_{31} $. Also note that these conditions are easily satisfied when the colonization rates $r_{21}$ and $r_{31}$ are negligibly small, which is the case for most of the parameter phase space, whereas Eq.~\eqref{MTEapprox2} breaks down only in non-WKB parameter regimes, see below.

\begin{figure}[t!]
	\includegraphics[width=1\linewidth]{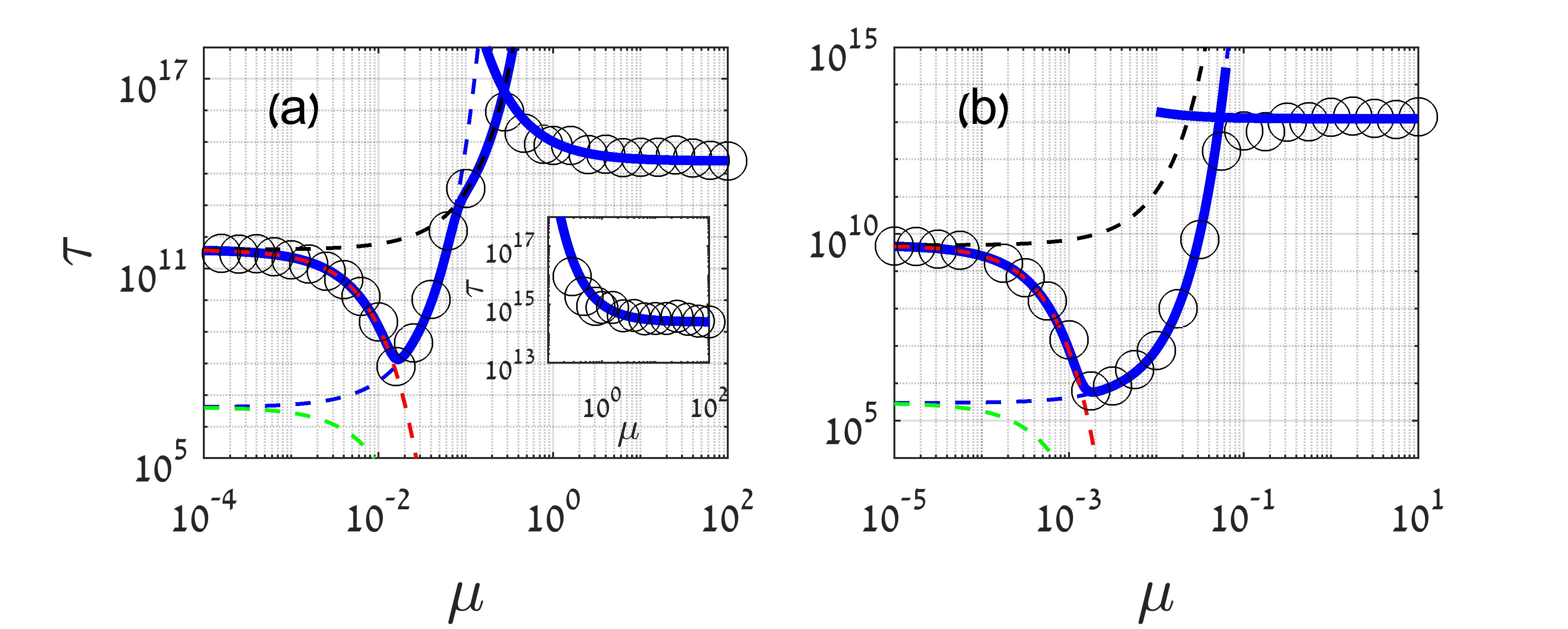}
	\caption{ MTE as a function of $\mu$ as given by Eq.~(\ref{MTEgeneralSM}) for slow migration, and by Eq.~(\ref{SwithCorrection}) for fast migration (solid lines), compared with WE simulations (circles). The dotted lines are the individual transition times, black for $r^{-1}_{13}$, red for $r^{-1}_{24}$, blue for $r^{-1}_{12}$ and green for $r^{-1}_{34}$, see text. In (a) parameters are $N_1 = 2300$, $N_2 = 2000$, $\delta_1 = 0.25$, $\delta_2 = 0.21$, and $\alpha = 1$. In the inset we show the fast migration result. In (b) $N_1 = N_2 = 10^5 $, $ \delta_1 = 0.065$, $ \delta_2 = 0.05$ and $\alpha =1$. }
	\label{fig2sm}
\end{figure}

In light of the arguments specified above, the MTE is given by the following generic expression:
\begin{equation} \label{MTEgeneralSM}
\tau  \simeq  \min\!\left\{\!\max\!\left[\frac{1}{r_{12}},\frac{1}{r_{24}}, \frac {r_ {21}} {r_ {12} r_ {24}} \right]\!, \max\!\left[\frac{1}{r_{13}},\frac{1}{r_{34}}, \frac {r_ {31}} {r_ {13} r_ {34}} \right]\!\right\}\!,
\end{equation}
see Eq.~(\ref{MTEgeneral}) in the main text. In Figs.~\ref{fig2sm} and \ref{fig3sm} we plot the MTE as a function of $\mu$ (see also Fig.~3 in the main text). Here, we plot the individual switching times ($r^{-1}_{ij}$), see next subsection, showing how the MTE depends, for slow migration, on the individual rates.

\subsection{Slow migration - analytical rates}
In this subsection we compute the rates $r_{ij}$ in the limit of $\mu\ll 1$. We start by showing that in this limit, extinction occurs in a serial manner with an overwhelming probability; this is done by explicitly calculating $r_{14}$. In the leading order in $\mu$, the action is the sum of all independent actions [11]
\begin{eqnarray}
S_{14} = \sum_{i = 1}^2 \int_{(\bar{\bm{x}}_{+, +})_i}^{(\bar{\bm{x}}_{-, -})_i} p_i dx_i \simeq S_0(\delta_1) + \kappa S_0(\delta_2).
\end{eqnarray}
Since in the limit  $\mu\to 0$, this expression coincides with the sum of the actions $S_{12}$ and $S_{24}$ or $S_{13}$ and $S_{34}$, we find that $r_{14} \sim r_{12} r_{24} \sim r_{13} r_{34}$, which means that $r_{14}$ is exponentially smaller than any of the individual rates. Clearly, the same argument also applies for rates $r_{23}$ and $r_{32}$ which include transitions resulting in changes in both patches. Moreover, the fact that $r_{14}$, $r_{23}$ and $r_{32}$, are negligibly small compared to all other rates, was also verified by WE simulations.
Note, that similarly as done in Ref. [11], subleading-order corrections in $\mu\ll 1$ can also be computed for $r_{14}$. Yet, we do not give the corrections here as $r_{14}$ itself is negligible in the slow migration regime.

We now compute $r_{13}$ -- the extinction rate of patch 1 while patch 2 remains colonized; the rest of the rates can be computed in a similar manner. Our starting point is Hamiltonian~\eqref{HamiltonianLowMigration} in the main text; this effective 1D Hamiltonian is obtained by assuming that $x_2$ and $p_2$ fluctuate around their FP with demographic noise of order $(\kappa N)^{-1/2}$, and additional migrational noise of order $\mu$, where both $\mu$ and $(\kappa N)^{-1/2}$ are small and uncorrelated, see main text. As a result, the optimal path -- the zero-energy trajectory of Hamiltonian~\eqref{HamiltonianLowMigration} in the main text -- reads
\begin{equation} \label{optimal_path_low_migrationNotApprox}
p_1 (x_1)= \log \left(\frac{d_1(x_1)+\mu x_1}{b_1(x_1)+\mu \alpha \kappa (1+\delta_2) }\right),
\end{equation}
which, for $\mu\ll 1$, can be approximated as
\begin{equation} \label{optimal_path_low_migration}
p_1(x_1) = \log \left[\frac{d_1(x_1)}{b_1(x_1)}\right]  \!+\! \mu \left[\frac{x_1 }{d_1(x_1)} \!-\!  \frac{ \alpha \kappa (1 + \delta_2)}{b_1(x_1)}\right].
\end{equation}
Therefore, the action satisfies
\begin{equation} \label{MeanSwitch}
S_{13} =  \sum_{k = 1}^2 \int_{(\bar{\bm{x}}_{+, +})_k}^{(\bar{\bm{x}}_{-, +})_k} p_k(x_k) dx_k \simeq \int_{(\bar{\bm{x}}_{+, +})_1 }^{(\bar{\bm{x}}_{-, +})_1} p_1(x_1) dx_1,
\end{equation}
where the integral limits are given by Eqs.~(\ref{x_mf_lowMigration}), and the integral over $p_2(x_2)$ was ignored as it contributes only $\OO(\mu^2)$ terms. Performing the integral in Eq.~\eqref{MeanSwitch} with $p_1(x_1)$ given by~\eqref{optimal_path_low_migration}, we find the action $S_{13}$ up to first order in $\mu$. Note that, computing the actions $S_{12}$, $S_{24}$ and $S_{34}$ corresponding to the transitions between FP1 and FP2, FP2 and FP4 and FP3 and FP4, respectively (see Fig.~2 in the main text), can be done in a similar manner, which yield Eqs.~(\ref{Action_rij}) of main text. Finally, given these actions, the transition rates are given by $r_{ij} = e^{-N S_{ij}}$ for $ij = \{12, 13, 34, 24\}$.

\begin{figure}[t!]
	\includegraphics[width=1\linewidth]{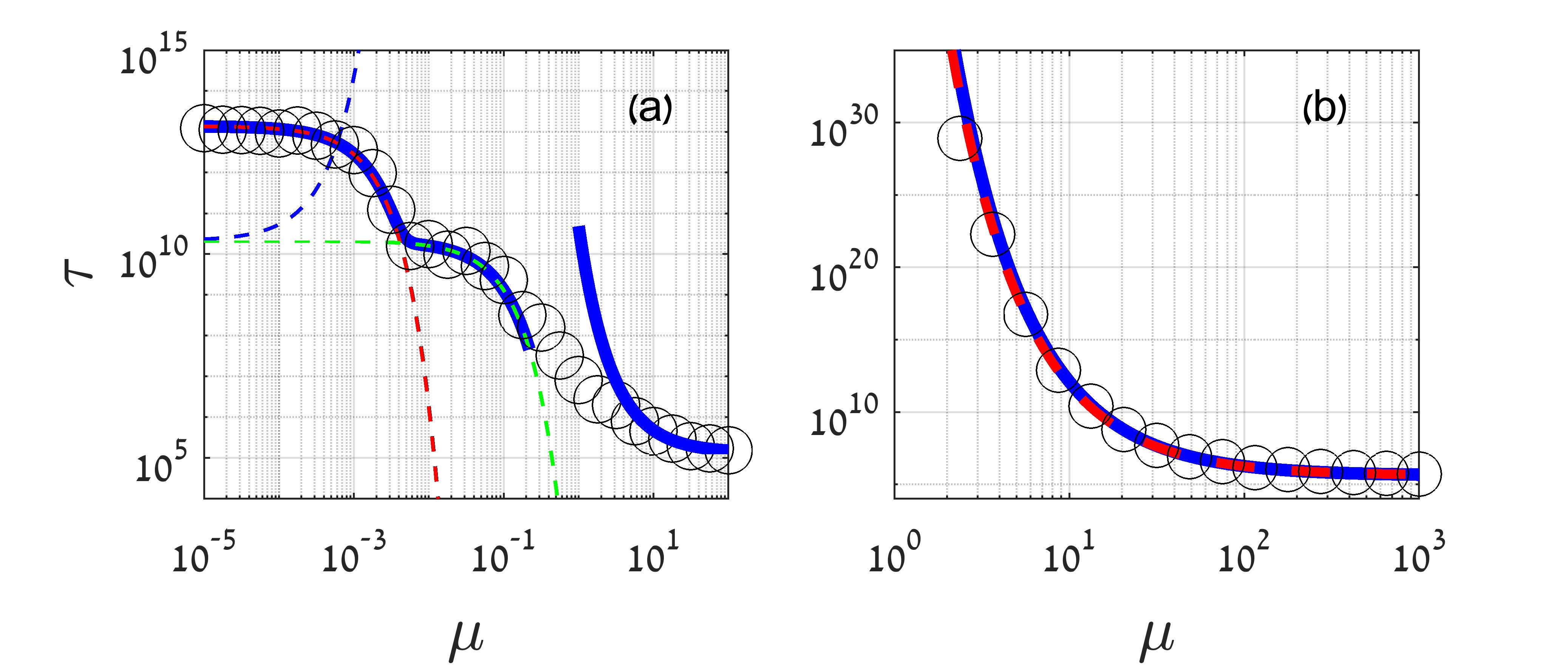}
	\caption{(a) MTE as a function of $\mu$ as given by Eq.~(\ref{MTEgeneralSM}) for slow  migration, and Eq.~(\ref{SwithCorrection}) for fast migration (solid lines), compared with WE simulations (circles). The dotted lines are the individual transition times, black for $r^{-1}_{13}$, red for $r^{-1}_{24}$ (overlaps with black line), blue for $r^{-1}_{12}$ and green for $r^{-1}_{34}$, see text. Parameters are $N_1 = 10^4$, $ N_2 = 50$, $ \delta_1 = 0.16$, $ \delta_2 = 0.8$ and $ \alpha = 1$. (b) MTE as a function of $\mu$ as given by Eq.~(\ref{SwithCorrection}) for fast migration (solid line), compared to the approximation for $\kappa \ll 1$, Eqs.~(\ref{ActionSmallKappaZeroOrder}) and~(\ref{ActionSmallKappaFirstOrder}) (dashed line). Parameters are $N_1 = 10^5$, $ N_2 = 800$, $  \delta_1 = 0.4$, $ \delta_2 = 0.72$ and $ \alpha = 1$.}
	\label{fig3sm}
\end{figure}

In Figs.~\ref{fig2sm} and~\ref{fig3sm}(a) we plot, alongside $\tau$, the individual mean transition times, corresponding to $r_{12}^{-1}, r_{13}^{-1}, r_{24}^{-1}, r_{34}^{-1}$, see figure captions. In this way we demonstrate the dependence of $\tau$ on these transition times, as given by the maximum functions in Eq.~\eqref{MTEgeneralSM}. Indeed, in Fig. \ref{fig2sm}, as $\mu$ increases the system switches between the decreasing $r_{24}^{-1}$ and the increasing $r_{12}^{-1}$, while in Fig.~\ref{fig3sm}(a) the system switches between the decreasing $r_{24}^{-1}$ and $r_{13}^{-1}$ (since here $\kappa \ll 1$, these rates are almost indistinguishable) and the also decreasing $r_{34}^{-1}$. In both figures the parameters were chosen such that recolonization is highly improbable.

\begin{figure}[t]
	\includegraphics[width=1\linewidth]{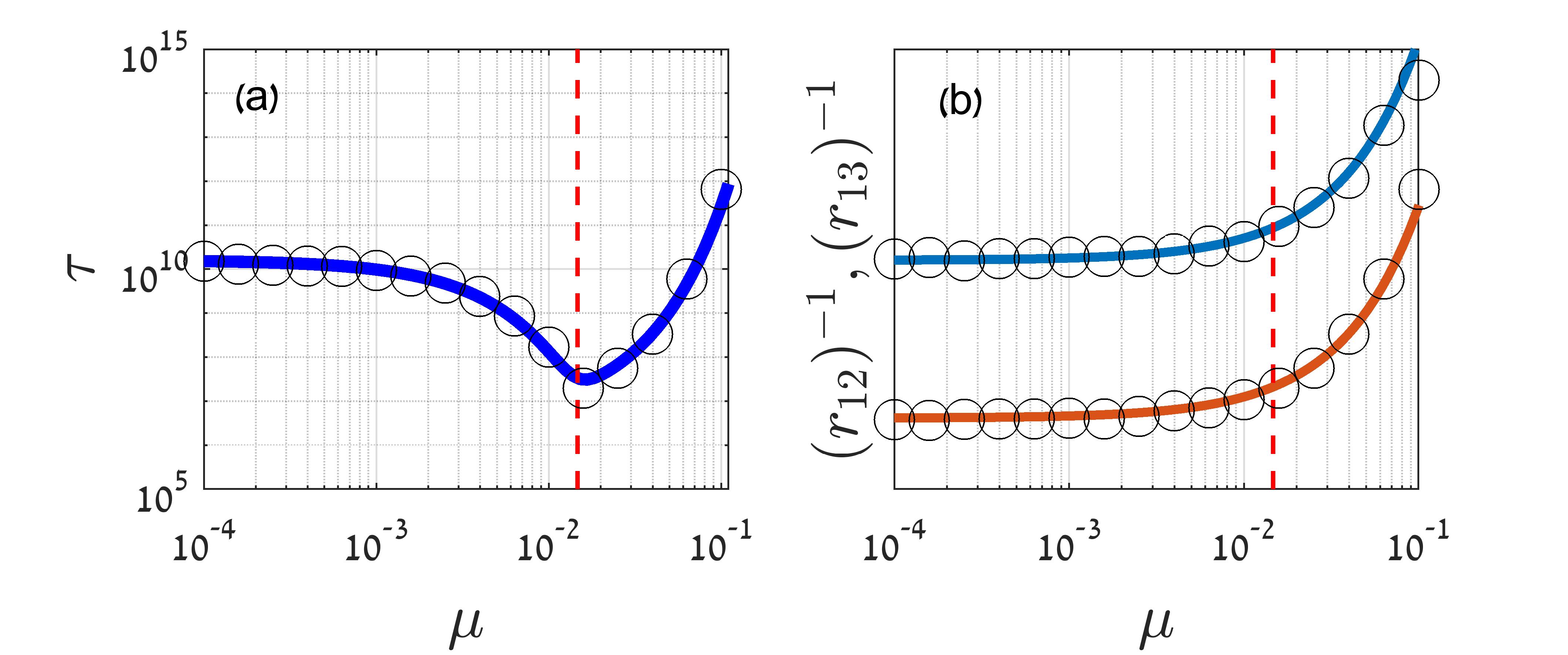}
	\caption{  (a) MTE as a function of $\mu$ as given by Eq.~(\ref{MTEgeneralSM}). (b) Extinction rate of a single patch as a function of $\mu$, where the second patch is colonized [see Eqs.~(\ref{Action_rij}) of main text]. Both panels are compared with WE simulations (circles), and we denote $\mu_{crit}$, Eq.~\eqref{mu_min1}, by a vertical dashed line. Parameters are  $N_1 = N_2 = 2000 $, $ \delta_1 = 0.25 $, $ \delta_2 = 0.21$ and $ \alpha = 1$.}
	\label{fig4sm}
\end{figure}

Having computed the individual extinction rates, we compare in Fig.~\ref{fig4sm} the extinction of local patches with global extinction of the metapopulation. This figure demonstrates that while each patch locally profits from increasing migration, the MTE of the global metapopulation decreases with increasing $\mu$. These results support our claim in the main text, that when the local dynamics include the Allee effect, even though each patch separately experiences a decrease in its local extinction risk (``rescue effect"), the  extinction risk of the entire metapopulation increases.

In addition to the extinction rates, colonization rates $r_{21}$ and $r_{31}$, can also be obtained, by integrating over optimal path~\eqref{optimal_path_low_migration} between the corresponding FPs. For example, colonization rate of patch 1 while patch 2 is colonized (transition from FP3 to FP1) is found by integrating between $(\bar{\bm{x}}_{0, +})_1$ and $(\bar{\bm{x}}_{-,+})_1$, given by Eqs.~(\ref{x_mf_lowMigration}). Yet, due to boundary issues, see below, one should take care in approximating the optimal path in $\mu$. To circumvent this problem, we first integrate over optimal path~\eqref{optimal_path_low_migrationNotApprox} and then approximate the result up to $\OO(\mu)$. A similar method is used to compute the transition from FP2 to FP1. This results in
\begin{eqnarray} \label{S_31}
&S_{31} =  \left(\delta_1+2 (1-\delta_1^2)^{1/2} \arcsin
  \left[\left(\frac{1-\delta_1}{2}\right)^{1/2}\right]-1\right) \nonumber  \\&- \mu^{1/2}\frac{\pi [\kappa \alpha(1+\delta_2) (1-\delta_1^2)]^{1/2}}{2^{1/2}}
   \\&+\mu \left((1-\delta_1^2)^{1/2} \arcsin\left[\left(\frac{1-\delta_1}{2}\right)^{1/2}\right]+\frac{\kappa \alpha(1+\delta_2) \left(\delta_1+3\right)}{2} \right), \nonumber
\end{eqnarray}
and
\begin{eqnarray} \label{S_21}
&S_{21} = \kappa \left(\delta_2+2 (1-\delta_2^2)^{1/2} \arcsin
  \left[\left(\frac{1-\delta_2}{2}\right)^{1/2}\right]-1\right)  \nonumber\\&- \mu^{1/2}\frac{\pi ((1+\delta_1) \left(1-\delta_2^2\right) \kappa)^{1/2}}{2^{1/2}}
   \\&+\mu \left(\!\alpha(1-\delta_2^2)^{1/2} \kappa \arcsin\left[\left( \frac{1-\delta_2}{2}\right)^{1/2}\right]+\frac{ (1+\delta_1)(\delta_2+3)}{2}\right). \nonumber
\end{eqnarray}
The colonization rates are thus given by $r_{21} = \exp(-N S_{21})$ and $r_{31} = \exp(-N S_{31})$. Note, that Eqs.~\eqref{S_31} and~\eqref{S_21} contain a $\OO(\mu^{1/2})$ term. This occurs due to the boundary issues previously mentioned; when integrating over the optimal path, since the lower boundary scales with $\mu$, it can be checked that the contribution from the integral over the optimal path, up to $x=\OO(\mu^{1/2})$, is not negligible.

Finally, having found all the extinction and colonization rates, we have numerically confirmed that for a reasonable choice of parameters, as long as $\delta_i$ is not too close to $1$, the rate of colonization of patch $i$ is negligible compared to its extinction rate (given that the other patch is colonized).

\subsection{Critical migration rate}
\begin{figure}[t!]
	\includegraphics[width=1.0\linewidth]{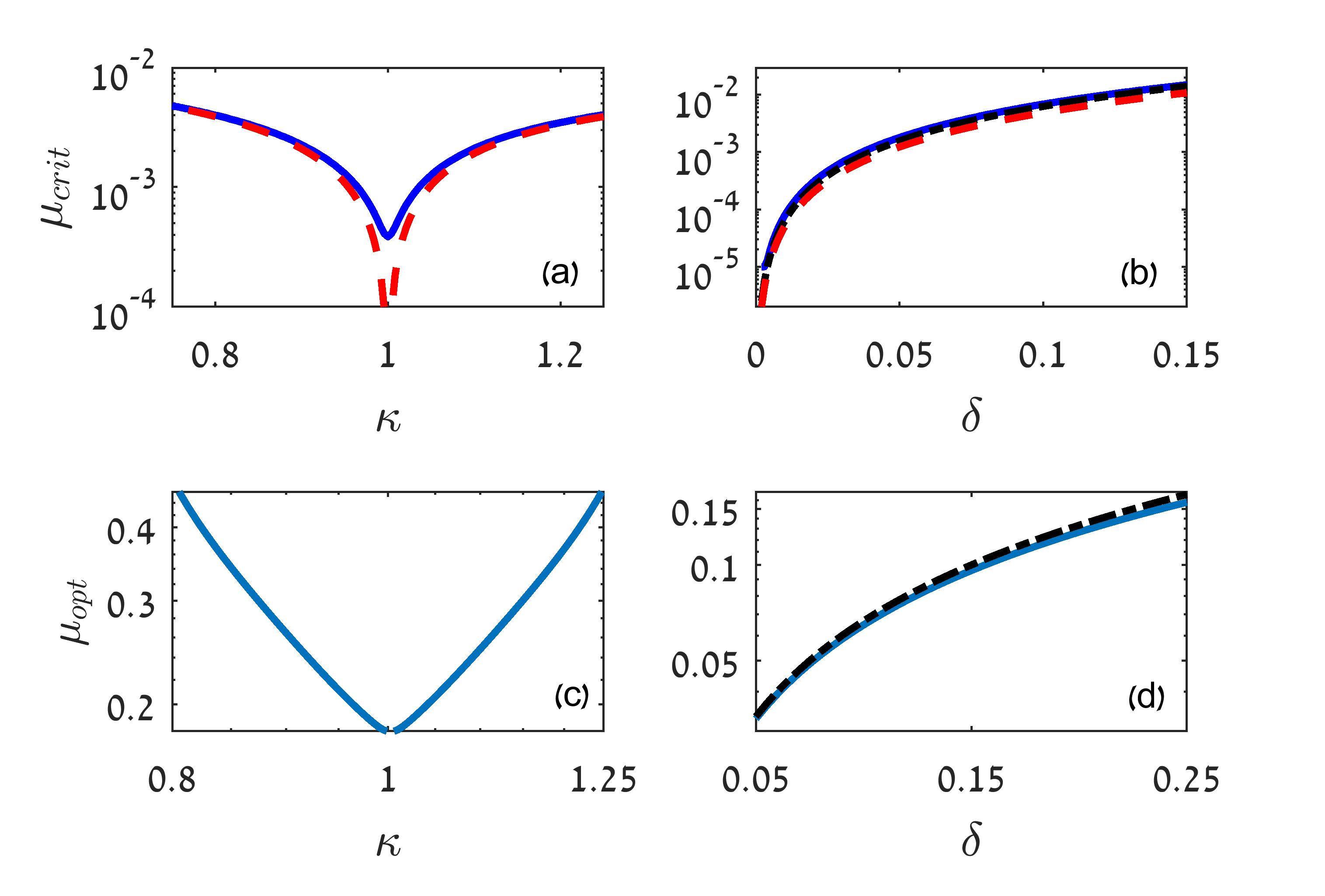}
	\caption{Critical and optimal migration rates. In (a) we plot $\mu_{crit}$ versus $\kappa$ as predicted by the exact form of $\tau$ given by Eq.~(\ref{MTEexact}) (solid line) and as given by Eq.~(\ref{mu_min1}) (dashed line). Parameters are $\delta_1 = \delta_2 = 0.2, \alpha = 1$. In (b) we plot the $\mu_{crit}$ as a function of $\delta \equiv \delta_1 = 2 \delta_2$. The solid line is the exact result as obtained from Eq.~\eqref{MTEexact}, the dashed line is Eq.~(\ref{mu_min1}) and the dashed-dotted line is the bifurcation result [Eq.~\eqref{mu_crit_bif}]. Parameters are $\kappa = 1/\alpha = 1/2$.  In (c) we plot $\mu_{opt}$ given by numerically solving Eq.~\eqref{GeneralEqForMuOpt}; here we could not find a simple analytical expression equivalent to Eq.~\eqref{mu_min1}. Parameters are $\delta_1 = \delta_2 = 0.3, \alpha = 1$. In (d) we plot $\mu_{opt}$ given by numerically solving Eq.~\eqref{GeneralEqForMuOpt} (solid line) along with the result close to the bifurcation limit (dashed-dotted line), see text. Here $\delta_1 = \delta_2 = \delta, \kappa = 1/\alpha = 7/11$.}
	\label{fig5sm}
\end{figure}

In this subsection we compute the value of $\mu_{crit}$, as well as the conditions for which it is a global minimum at a finite migration rate. These can be done numerically using the exact form of the MTE, Eq.~(\ref{MTEexact}), see \textit{e.g.}, Fig.~\ref{fig5sm}(a-b). However for all practical purposes the approximated MTE, Eq.~(\ref{MTEgeneralSM}), is sufficient.

For simplicity, here we find the critical migration rate in the case of negligible colonization rates. In this case, the maximum functions of Eq.~(\ref{MTEgeneralSM}) switch between $r^{-1}_{24}$ and $r^{-1}_{12}$ or between $r^{-1}_{34}$ and $r^{-1}_{13}$. The critical migration rate is thus given by equating $S_{24} = S_{12}$ or $S_{34} = S_{13}$:
\begin{equation} \label{mu_min1}
\mu_{crit} = \frac{S_0\left(\delta _1\right)-\kappa  S_0\left(\delta _2\right)}{ \delta _1 - S_0(\delta_1)/2 + C -\alpha  \kappa   \left[\delta _2- S_0(\delta_2)/2 \right]}.
\end{equation}
Here the value of $C$ depends on the relative stability of the patches. If $S_0(\delta_1 ) > \kappa S_0(\delta_2 )$, then $C = \delta_2\left(\delta _1+1\right)$ and the migration rate is indeed a minimum as long as the correction in $S_{12}$ is positive, \textit{i.e.}, $\delta_2\left(\delta _1+1\right) >\alpha  \kappa  \delta _2- S_0(\delta_2)/2$. On the other hand, if $S_0(\delta_1 ) < \kappa S_0(\delta_1 )$, then $C = \alpha \kappa \delta_1 (\delta_2+1)$ and this is a minimum as long as the correction in $S_{13}$ is positive, \textit{i.e.} $ \alpha \kappa \delta_1 (\delta_2+1) > \delta _1- S_0(\delta_1)/2$.

Equation~(\ref{mu_min1}) drastically simplifies close to the bifurcation limit where $\delta_1,\delta_2\ll 1$. Taking $\delta_1 \equiv \delta$ and $\delta_2 = u \delta$, where $u =\OO(1)$, Eq.~\eqref{mu_min1} simplifies to:
\begin{equation} \label{mu_crit_bif}
\mu_{crit} \simeq \max\left\{\frac{2}{3} \delta^2 \left(1 - \kappa u^3\right),\frac{2}{3 u} \delta^2 \left(\kappa  u^3-1\right)\right\},
\end{equation}
where we note that close to the bifurcation limit, since the colonization threshold is very close to the colonized state, the colonization rates are generally negligible.

Two examples of the existence of a global minimum at a finite $\mu$ are given in Fig.~\ref{fig2sm}, while in Fig.~\ref{fig3sm} we provide a counter example where the global minimum is obtained at $\mu\to\infty$. In Fig.~\ref{fig5sm}(a-b) we plot $\mu_{crit}$ as a function of $\delta$ and $\kappa$. In Fig.~\ref{fig5sm}(a) we compare Eq.~\eqref{mu_min1} with the critical migration rate obtained numerically by finding the minimum of the exact form of $\tau$, given by Eq.~\eqref{MTEexact}.  We find that Eq.~(\ref{mu_min1}) is a good approximation as long as $\kappa$ is not too close to $1$; indeed when $\kappa$ approaches $1$, the assumption that the transition rates exponentially differ from each other breaks down, which invalidates Eq.~\eqref{MTEgeneralSM} and correspondingly, Eq.~\eqref{mu_min1}. In Fig.~\ref{fig5sm}(b) we show that Eq.~(\ref{mu_crit_bif}) is a good approximation to Eq.~(\ref{mu_min1}) close to the bifurcation limit.

\subsection{Fast migration}
In this subsection we derive $H_{\text{fast}}^{(0)}$ and $S_{\text{fast}}^{(0)}$, Eqs.~(\ref{HamiltonianHighMigrationDdim}) and (\ref{ActionHighMigration2dim}), in the main text.
To this end we use a similar method to that presented in Ref. [12], but go beyond their leading-order result and compute subleading corrections in $\zeta=\mu^{-1}$ as well.
Our starting point is Hamiltonian~\eqref{HamiltonianDdim} in the main text. Assuming a fast migration rate, $\mu \gg 1$, we rescale the Hamiltonian using $\zeta$:
\begin{equation} \label{HamiltonianDdimH0H1}
H(\bm{x}, \bm{p})=H_{0}(\bm{x}, \bm{p})+\zeta H_{1}(\bm{x}, \bm{p}),
\end{equation}
with
\begin{eqnarray} \label{Hamiltonian2DimH0}
&&H_0(\bm{x}, \bm{p} ) = x_1 \left(e^{p_2 - p_1} - 1\right)+ x_2\alpha \left(e^{p_1 - p_2} - 1\right), \\
&&H_{1}(\bm{x}, \bm{p})=\sum_{i =1}^2 \left\{\left(e^{p_{i}}-1\right) \left[b_i(x_i)-e^{-p_i} d_i(x_i)\right] \right\}. \nonumber
\end{eqnarray}
That is, the local dynamics act as a perturbation to $H_0$, which includes the migration terms. Assuming that the total population size varies much slower than that of each individual patch, we use the following transformation
\begin{eqnarray} \label{TransformationHighMigration}
Q\!=\!  x_1 + x_2 \;, \;
q\!=\!x_2 \;, \;
P\!=\!\frac{1}{2} (p_1 + p_2) \;,\;
p\!=\!p_2 .
\end{eqnarray}
This transformation is motivated by the requirement that both $Q$ and $P$ are slowly varying variables compared to $q$ and $p$.
We now substitute this transformation into Hamiltonian~(\ref{HamiltonianDdimH0H1}) and write down the Hamilton equations, $\dot{p} =  -\partial_q H(Q, q, P, p) $ and $\dot{q} = \partial_p H(Q, q, P, p)$. Putting $\dot{p}=\dot{q}=0$, \textit{i.e.}, assuming the fast variables instantaneously equilibrate to some $(Q,P)$-dependent functions, and solving the resulting algebraic equations for $q$ and $p$ perturbatively with respect to $\zeta$, we find
\begin{equation} \label{ScalingFastVar}
q = \frac{Q}{1 + \alpha} + \zeta q^{(1)}(Q,P) \;\;,  \;\;\;p = P + \zeta p^{(1)}(Q,P) ,
\end{equation}
where $q^{(1)}(Q,P)$ and $p^{(1)}(Q,P)$ are (known) functions of $Q$ and $P$.
Substituting $q$ and $p$ from Eq.~(\ref{ScalingFastVar}) into Hamiltonian~(\ref{HamiltonianDdimH0H1}), keeping terms up to sub-leading order in $\zeta$, and dividing the result by $\zeta$ we arrive at an approximation for the  Hamiltonian in the fast migration regime:
\begin{equation} \label{H_slow_with_correction}
H_{\text{eff}} = H^{(0)}_{\text{eff}}(Q, P) + \zeta H_{\text{eff}}^{(1)}(Q, P) +\OO(\zeta^2).
\end{equation}
Here
\begin{equation} \label{HamiltonianHighMigrationDdimSM}
H_{\text{eff}}^{(0)}(Q, P) = \left(e^{P}-1\right) \left[\tilde{b}(Q) -e^{-P} \tilde{d}(Q)\right], \nonumber
\end{equation}
$\tilde{b}(x) = 2 x^2/[\tilde{\kappa}(1+1/\alpha)(1 - \tilde{\delta}^2)]$,  $\tilde{d}(x) =  x + x^3/[\tilde{\kappa}^2(1+1/\alpha)^2(1 - \tilde{\delta}^2)]$, while $H_{\text{eff}}^{(1)}(Q, P) $ is a (known) function of $Q$ and $P$, but too long to be explicitly presented.  Equating Hamiltonian~\eqref{H_slow_with_correction} to zero, yields
\begin{equation}\label{optpathfast}
P = P^{(0)}+ \zeta P^{(1)},\;\;\;\;P^{(0)} = \ln[\tilde{d}(Q)/\tilde{b}(Q)],
\end{equation}
while $P^{(1)}$ is a (known) function of $Q$, but too long to be explicitly presented.
Finally, since our transformation of variables is not canonical, the action can be written by using Eqs.~(\ref{TransformationHighMigration}) and ~\eqref{ScalingFastVar}:
\begin{eqnarray} \label{SwithCorrection}
&&S_{\text{fast}} =\int_{\xi_{+}+\OO(\zeta)}^{\xi_{-}+\OO(\zeta)}  \! p_1 dx_1 +  \int_{\xi_{+}+\OO(\zeta)}^{\xi_{-}+\OO(\zeta)}  \!p_2 dx_2
\\&& \simeq  \int_{Q_{+}+\OO(\zeta)}^{Q_{-}+\OO(\zeta)}\! P^{(0)}dQ   + \zeta\int_{Q_{+}}^{Q_{-}}\!\!\left( P^{(1)} + \frac{1-\alpha  }{1+\alpha}p^{(1)}\right)dQ . \nonumber
\end{eqnarray}
Here, $Q_{\pm} = (1 + 1/\alpha)\xi_{\pm}$ is the combined population size of the two patches for the colonized state ($Q_+$) and for the colonization threshold ($Q_-$), and $\xi_\pm$ are given by Eq.~(\ref{mf_high_migration}). The approximation in the second line of Eq.~\eqref{SwithCorrection} contains two integrals: in the first we integrate over the zeroth-order trajectory, $P^{(0)}$, and take into account possible corrections to the limits, while in the second, we integrate over the corrections to the trajectory, but neglect corrections to the limits as these contribute to the action only terms which are $\OO(\zeta^2)$.
To find the second integrand, we note that $p^{(1)}$ is a known function of $Q$ and $P$, where the latter has to be evaluated at $P^{(0)}$, given by Eq.~(\ref{optpathfast}) (since higher order corrections in $P$ contribute only ${\cal O}(\mu^2)$ terms to the integral). As a result, since $P^{(1)}$ is also a known function of $Q$, in the fast migration limit these integrals can be calculated, for any set of parameters, which yields $S_{\text{fast}}$ up to subleading-order in $\zeta$. In the following we present explicit results for both the zeroth-order term of $S_{\text{fast}}$ and the ${\cal O}(\zeta)$ correction (in particular limits where the result is amenable).

The leading-order contribution to $S_{\text{fast}}$ reads:
\begin{equation} \label{ActionHighMigrationApprox}
S_{\text{fast}}^{(0)}  = \int_{Q_{+}}^{Q_{-}} P^{(0)} dQ = (1+1/\alpha) \tilde{\kappa } S_0(\tilde{\delta}),
\end{equation}
which coincides with Eq.~(\ref{ActionHighMigration2dim}) of main text. Note, that this result indicates that our transformation of variables~(\ref{TransformationHighMigration}) is canonical up to $\OO(\zeta)$.

\begin{figure}[t!]
	\centering
	\includegraphics[width=1.0\linewidth]{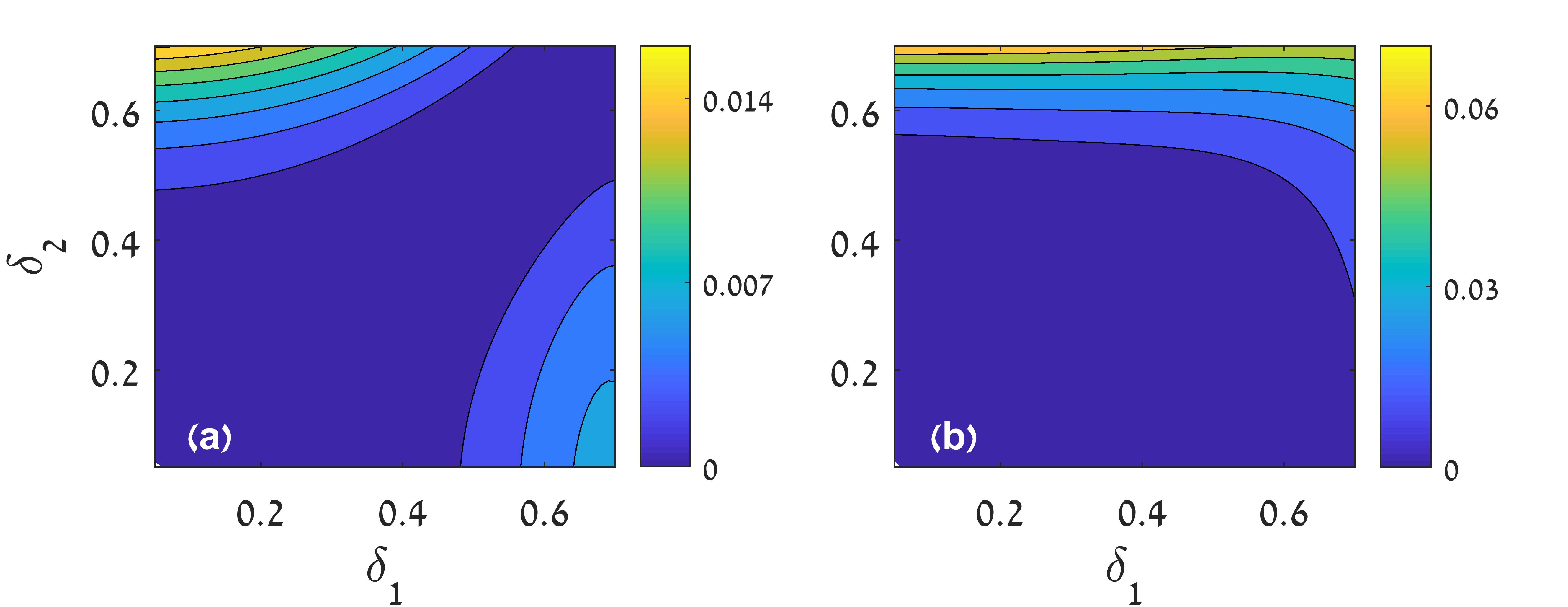}
	\caption{$S_{\text{fast}}^{(1)}$, the correction in $\zeta$ of Eq.~\eqref{SwithCorrection}, as a function of $\delta_1$ and $\delta_2$. Parameters are $\alpha = 2$ and in (a) $\kappa = 1/\alpha = 0.5$ while in (b) $\kappa = 1$. In (a), along the line $\delta_1 = \delta_2$, we find that $S_{fast}^{(1)} = 0$ . }
	\label{fig6sm}
\end{figure}

We now turn to discuss the correction term, which we term $S_{\text{fast}}^{(1)}$ such that $S_{\text{fast}} = S_{\text{fast}}^{(0)} + \zeta S_{\text{fast}}^{(1)}$. Although the full expression is cumbersome to give in full, we include this correction in Figs.~\ref{fig2sm} and~\ref{fig3sm}.  Additionally, in Fig.~\ref{fig6sm} we plot $S_{\text{fast}}^{(1)}$ as a function of $\delta_1$ and $\delta_2$ for two different cases. We find that in most realistic cases $S_{\text{fast}}^{(1)}$ is positive, which suggests that in general there exists an optimal migration rate, which maximizes the metapopulation's lifetime, see below.

We now present two cases in which the correction $S_{\text{fast}}^{(1)}$ drastically simplifies. The first and simplest case, is of well mixed patches $\kappa = 1/\alpha $ and $ \delta_1 = \delta_2$. Here we find that the correction is zero, see \textit{e.g.}, Fig.~\ref{fig6sm}(a). This occurs since all terms that depend on $\zeta$ in Eq.~\eqref{SwithCorrection}, $P^{(1)} $, $ p^{(1)}$ and the corrections to the integral limits, are all equal to zero in this special case. Thus, for well mixed patches [12] the correction in $\zeta$ is zero, entailing that the survival probability reaches a steady value in the fast migration regime.

The second case in which a simplified expression for the correction can be found, is the limit of very different carrying capacities, $\kappa \ll 1$. This is an extreme scenario in which a loss of synchrony may lead to a ''source sink" dynamics, where the large patch (\textit{i.e.}, the patch with the large carrying capacity) becomes a sink to the neighboring small patch (\textit{i.e.}, with smaller carrying capacity). Above, we have shown that at the deterministic level, the smaller patch dictates the typical size of the population. Accounting for demographic noise, we find that for $\kappa\ll 1$, the leading-order action [Eq.~(\ref{ActionHighMigrationApprox})] becomes
\begin{equation} \label{ActionSmallKappaZeroOrder}
S_{\text{fast}}^{(0)} \simeq (\alpha +1) \kappa S_0(\Delta_2),
\end{equation}
where $S_0(\delta)$ is the action of an isolate patch as given in main text and $\Delta_2 = (\delta_2 ^2 - \alpha(1- \delta_2 ^2))^{1/2}$, as previously defined. Furthermore, in this limit, we can also compute the subleading-order correction to the action, which yields
\begin{equation} \label{ActionSmallKappaFirstOrder}
S_{\text{fast}}^{(1)} \simeq \alpha \kappa \left( 4 \text{arctanh}(\Delta_2) - 3 \Delta_2 -9S_0(\Delta_2)/2\right).
\end{equation}
These results give rise to a MTE that is independent of the details of the patch with the higher carrying capacity, \textit{i.e.}, $N$ and $\delta_1$. That is, for $\kappa\ll 1$, in addition to deterministically determining the metapopulation's mean, we find that the smaller patch also dictates the metapopulation's survival probability.

In Fig. \ref{fig3sm}(b), the analytical action given by Eqs.~\eqref{ActionSmallKappaZeroOrder} and~\eqref{ActionSmallKappaFirstOrder} is compared to the full action, given by Eq.~\eqref{SwithCorrection}, and very good agreement is observed.

\subsection{Optimal migration rate}
In this subsection we show how the optimal migration rate, denoted $\mu_{opt}$, for which the extinction risk is minimized, can be found. In general, $\mu_{opt}$ exists if $S_{\text{fast}}^{(0)} > \max\left\{S_0(\delta_1), \kappa S_0(\delta_2)\right\}$, and if the correction $S_{\text{fast}}^{(1)}$ is positive.

A general scaling for $\mu_{opt}$ can be found by comparing the solutions for slow and fast migration, \textit{i.e.}, by solving the following equation for $\mu_{opt}$:
\begin{equation} \label{GeneralEqForMuOpt}
\tau_{\text{slow}}(\mu_{opt}) = \tau_{\text{fast}}(\mu_{opt}),
\end{equation}
where the left hand side, $\tau_{\text{slow}}(\mu_{opt})$, is the slow-migration MTE given by Eq.~\eqref{MTEgeneralSM}, and the right hand side is the fast-migration MTE given by $\tau_{\text{fast}}(\mu_{opt}) = \exp\{N[S_{\text{fast}}^{(0)}+S_{\text{fast}}^{(1)}(\mu_{opt}^{-1})]\}$.
While the solution to Eq.~(\ref{GeneralEqForMuOpt}) can be found numerically, we now evaluate $\mu_{opt}$ close to the bifurcation limit where $\delta_1,\delta_2\ll 1$, and where Eqs.~\eqref{MTEgeneralSM} and ~\eqref{ActionHighMigrationApprox} drastically simplify. By further denoting $\delta_1 = \delta$ and assuming $\delta_2 = \delta u$ with $|u - 1| \ll 1$, we find a simple scaling law for the optimal migration rate: $\mu_{opt} \sim 2\delta/3 +\OO[\delta( u- 1)] $.

Examples of the numerical solution of Eq.~(\ref{GeneralEqForMuOpt}) for $\mu_{opt}$ as a function of $\kappa$, and a comparison between this solution and the result close to bifurcation, $\mu_{opt} \sim 2\delta/3$, as a function of $\delta$, are respectively given in Fig.~\ref{fig5sm}(c) and (d). In these examples the correction $S_{\text{fast}}^{(1)}$ is  positive such that there exists a maximum, apart from the special case of $\kappa = 1$ in Fig.~\ref{fig5sm}(c) where  $S_{fast}^{(1)}= 0$.

\section{Generalization to M patches}
In this section we provide generalization of our results to $M$ fully connected patches. This is done in both the slow and fast migration limits.
\subsection{The case of slow migration}
The actions describing the extinction of a single patch while the second patch experiences $\OO (\mu)$ changes [Eqs.~(\ref{Action_rij}) in the main text] can be generalized to $M$ connected patches. Here, we define $\phi^{in}_i = \sum_{j\neq i} \alpha_{ji} x_{j,s_j}^{(0)}$ as the incoming flux to patch $i$ given by the sum over all zeroth-order stable FPs of patches migrating into this patch [$s_j = (0, +)$]. Likewise we define $\phi^{out}_i = \sum_{j\neq i} \alpha_{ij}$ as the magnitude of the outgoing flux from patch $i$. Now, using similar arguments as in the two-patch case, yields:
\begin{equation} \label{ActionLowMigrationGeneral}
S_{i}(\phi^{in}_i) = \kappa_i S_0(\delta_i)(1-\mu \phi^{out}_i /2) + \mu \delta_i (\phi^{in}_i - \phi^{out}_i \kappa_i) ,
\end{equation}
where  $S_0(\delta_i)$ is given in main text. Since $\phi^{in}_i$ depends on $x_{j,s_j}^{(0)}$ for $j\neq i$ and $s_j = (0,+)$, it can take $2^{M - 1}$ different values, as each of the other patches can be either extinct or colonized. On the other hand, $\phi^{out}_i$ is determined only by patch $i$. The notation $S_{i}(\phi^{in}_i)$ is thus intended to emphasize that this action describes the extinction of patch $i$, given a specific influx $\phi^{in}_i$ (chosen out of $2^{M-1}$ possibilities) as dictated by the deterministic number of occupants in all other patches.

The colonization rate of patch $i$, given by Eqs.~(\ref{S_31}) and (\ref{S_21}), can also be generalized to M patches in a similar manner:
\begin{eqnarray} \label{ColonizationRateMpatches}
&\mathcal{S}_{i}(\phi^{in}_i) = \kappa _i \left(\delta _i\!+\!2 (1\!-\!\delta _i^2)^{1/2} \arcsin
 \left[ \left(\frac{1-\delta_i}{2}\right)^{1/2}\right]\!-\!1\right) \nonumber\\ &- \mu^{1/2}\frac{\pi (\phi^{in}_i \left(1-\delta _i^2\right) \kappa _i)^{1/2}}{2^{1/2}}
   \\&+\mu \left(\! \phi^{out}_i (1\!-\!\delta _i^2)^{1/2} \kappa _i \arcsin\left[\!\left(\frac{1-\delta _i}{2}\right)^{1/2}\!\right]+\frac{1}{2} \phi^{in}_i \left(\delta _i\!+\!3\right)\right) \!,\nonumber
\end{eqnarray}
where in the two-patch case $S_{31}$ and $S_{21}$ are obtained from this result for $i = 1, \phi^{out}_1 = 1, \phi^{in}_1 = \alpha\kappa(1 + \delta_2)$ and $i = 2, \phi^{out}_2 = \alpha, \phi^{in}_2 = (1 + \delta_1)$, respectively.

\subsection{Fast migration }
The zeroth-order result in the fast migration limit, Eq.~(\ref{ActionHighMigrationApprox}), can also be generalized to $M$ patches. Here we choose for simplicity $\mu_{ij}=\mu$. By conducting a similar calculation to the two-patch case one arrives at the following action:
\begin{equation} \label{ActionHighMigrationMdim}
S_{\text{fast}}^{(0)} = M \tilde{\kappa } S_0(\tilde{\delta}),
\end{equation}
where here $\tilde{\kappa}$ and $\tilde{\delta}$ are given by [12]
\begin{eqnarray}
&\frac{M}{\tilde{\kappa} (1-\tilde{\delta}^2)} =  \sum_{i =1}^M  \frac{1}{\kappa_i (1-\delta_i^2)} , \\
&\frac{M}{\tilde{\kappa}^2 (1-\tilde{\delta}^2)} =  \sum_{i =1}^M  \frac{1}{\kappa_i^2 (1-\delta_i^2)} , \nonumber
\end{eqnarray}
and it is required that $\tilde{\delta}$ be real, as in the two-patch case.

\section{Weighted Ensemble Simulations} \label{WEsimulations}

\begin{figure}[t!]
	\includegraphics[width=1\linewidth]{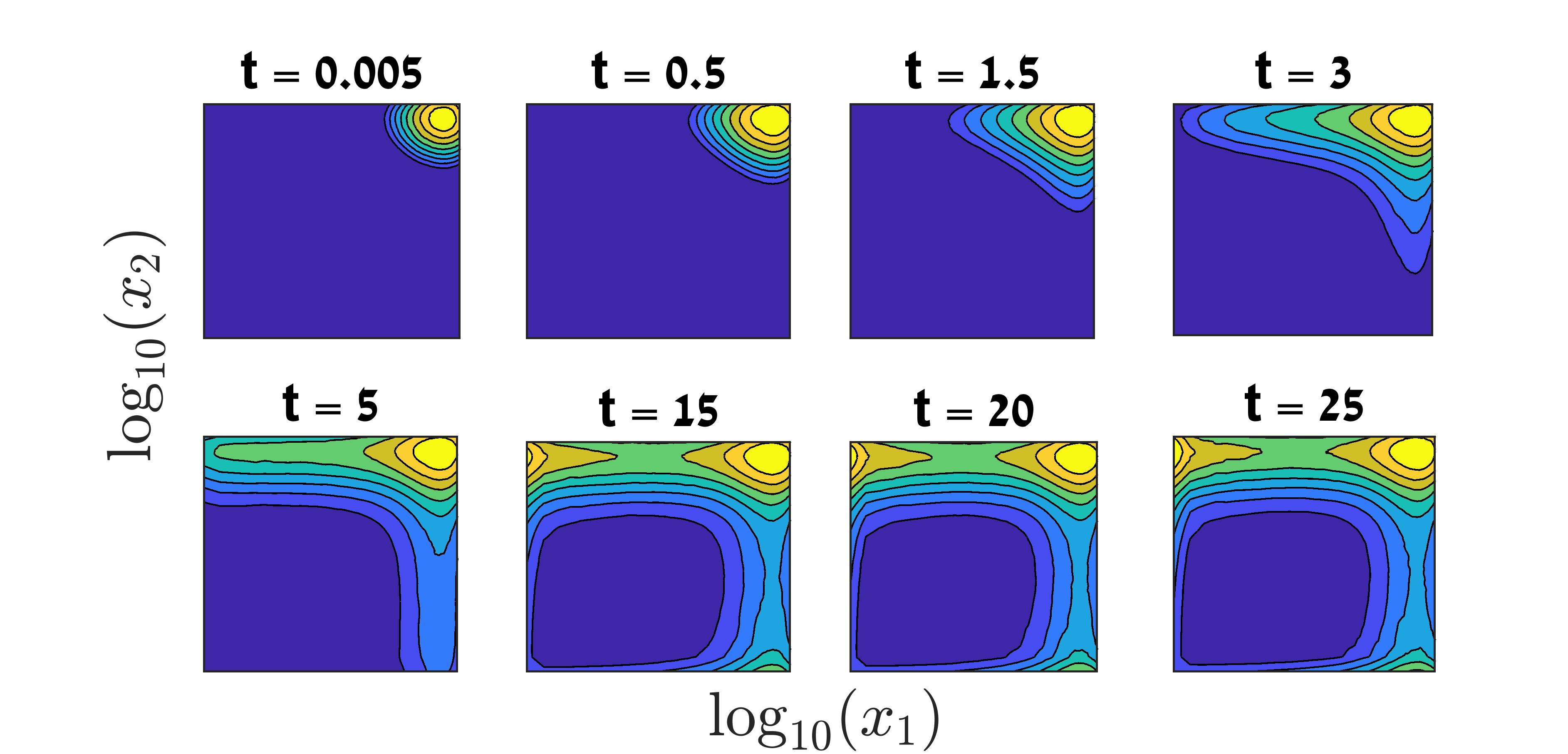}
	\caption{ Contour plot of the probabilities $\mathbb{P}_{n_1, n_2}$, see main text, obtained by WE simulations. Parameters  are $N_1 = N_2 = 100$, $ \delta_1 = 0.5$, $ \delta_2 = 0.4$, $ \mu = 10^-3$ and $\alpha = 1$.}
	\label{fig7sm}
\end{figure}

In this section we discuss the Weighted Ensemble (WE) simulations that were used to verify our analytic results. In our study, WE simulations were used to probe multiple transitions in space, and their associated probabilities. For example, in Fig.~\ref{fig7sm} we give snapshots from a WE simulation at different times. Here one observes the flow of probability from FP1 to both FP2 and FP3, and ultimately from FP2 and FP3 to FP4 [see Fig.~\ref{fig2sm}(a) in the main text], where at long times the system reaches a quasi-stationary distribution of population sizes. These simulations are used to verify all our analytical results, and many of our theoretical assumptions. For example, in the simulation depicted in Fig.~\ref{fig7sm} one observes that extinction occurs serially, and that transitions like FP1$\to$ FP4 occur with very low probabilities, see  above.

The basic idea of the algorithm we use is to run significantly more simulations in regions of interest, and to compensate for the bias, we distribute the weight of each trajectory accordingly. To this end, space is divided into bins, which can be predefined or interactively chosen (on the fly), to ensure sampling in specific regions of interest. We thus start the simulation with $m$ trajectories in proximity to a stable fixed point of the system. Each of the $m$ trajectories are given initial equal weights of $1/m$. The simulation consists of two general steps: (a) Trajectories are advanced in time for time $\tau_{WE}$, where the time-propagation method follows the Gillespie Algorithm [34, 35]; (b) Trajectories are re-sampled as to maintain $m$ trajectories in each occupied bin, while bins that are unoccupied remain so. An illustration of the method is given in Fig.~\ref{fig2}(b) in the main text, in which the number of bins is four and the number of trajectories is $m = 3$. The process of re-sampling itself can be done in various ways, as long as the distribution is maintained. In our simulation we used the original re-sampling method suggested by Huber and Kim [33].

\begin{figure}[t!]
\centering
	\includegraphics[width=0.7\linewidth]{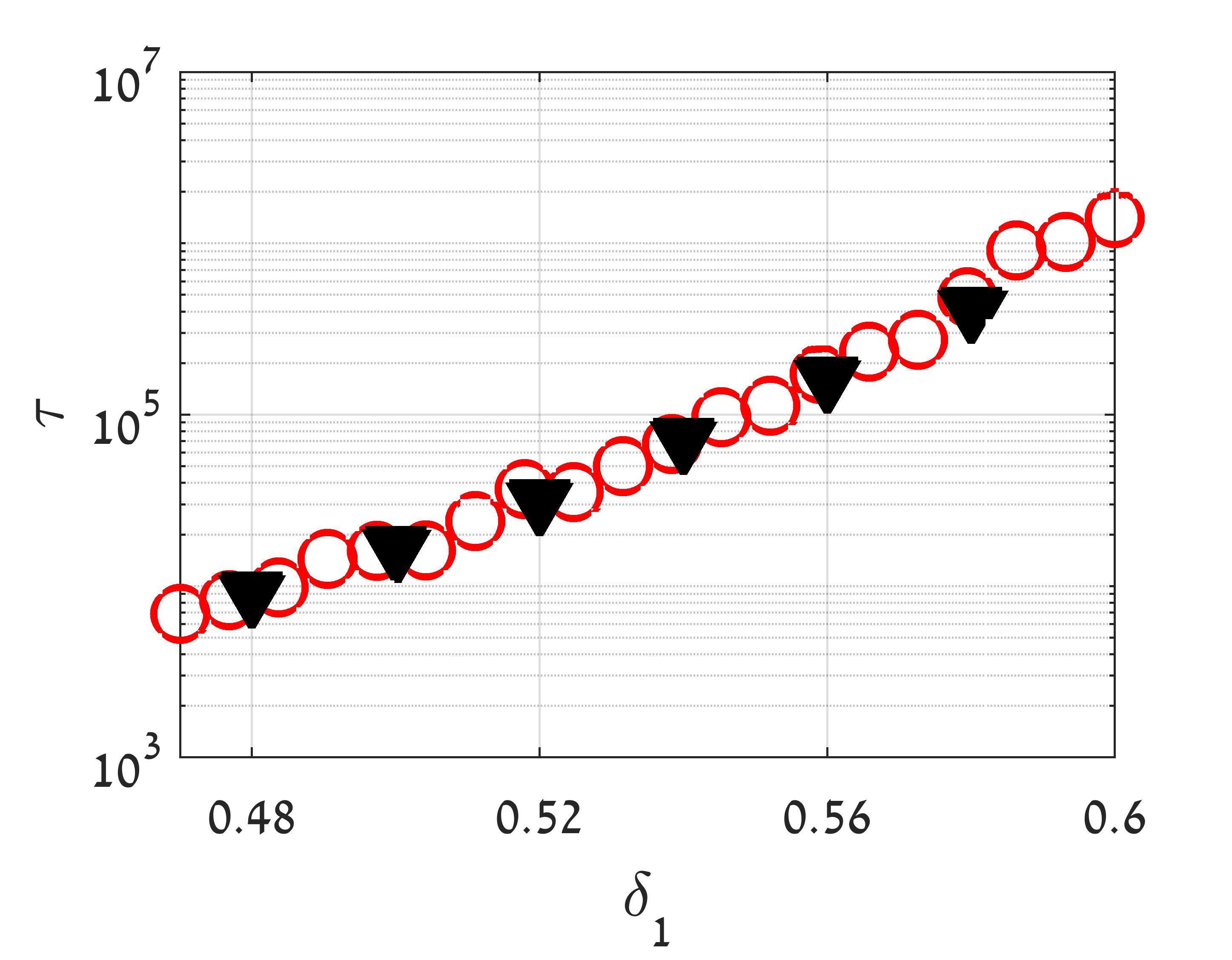}
\vspace{-2mm}
	\caption{MTE obtained by brute-force Monte-Carlo (full triangles) and WE (circles) simulations, as a function of $\delta_1$. Parameters are $N_1 = N_2 = 60$, $\delta_2 = 0.45$, $\mu =2$ and $\alpha = 1$. }
	\label{fig8sm}
\end{figure}

Note that $\tau_{WE}$ is chosen to be much shorter than the relaxation time of the system, $t_{r, +}^{relax}$, but much longer than the typical time between reactions, as to increase efficiency. We also stress that bins need to be chosen wisely: if chosen too far apart, trajectories will not reach remote regions, while if chosen close together the computational cost will be very high. Generally, there is a tradeoff between the number of bins and the trajectories per bin, assuming some memory limit. In our simulations, to achieve high efficiency we interactively changed the bins.

Error evaluation in our simulations was conducted numerically: by altering various parameters of the simulation, such as the number of bins and trajectories per bin, we were able to get an estimate of the error. In general we obtain a maximum error of $20\%$. This error is accounted for via the size of circles in all relevant figures.

Importantly, we checked that the results of the WE simulations coincide with brute force Monte-Carlo simulations in parameter regimes in which the latter are applicable, see \textit{e.g.}, Fig.~\ref{fig8sm}. We stress that WE simulations are much more efficient than brute-force Monte-Carlo simulations. The latter are very limited in probing rare events, as longer MTEs demand exponentially long simulation times, and they also lack the ability to easily separate different paths of extinction, which is at the center of this study. WE simulations are thus ideal for our purpose: longer MTEs do not require exponentially longer simulations and additionally, by measuring the flux between different meta-stable states, we can easily differentiate between global extinction and individual routes to extinction.

\section{Realistic Model}
\begin{figure}[h!]
\centering
\hspace{-6.2mm}\includegraphics[width=3.65in, height=1.5in, trim = {.6cm .6cm .6cm .6cm}, clip]{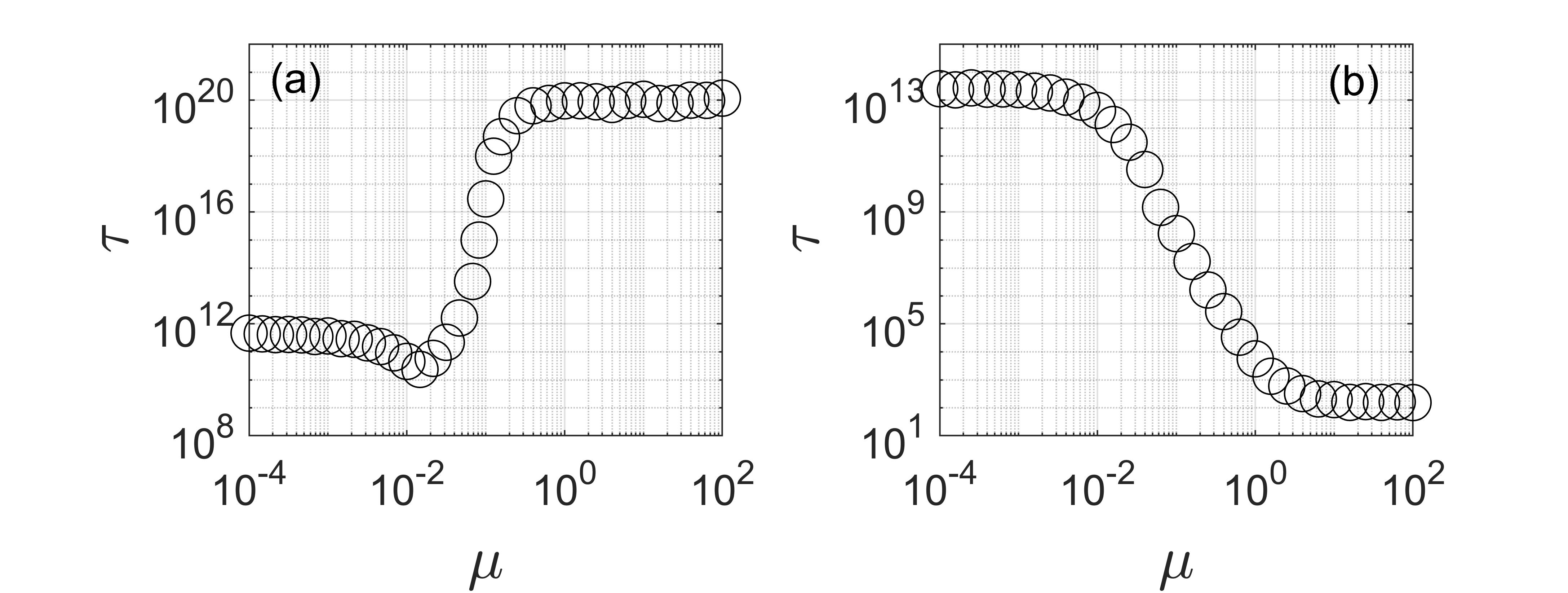}
\vspace{-2mm}
	\caption{MTE as a function of $\mu_{12} = \mu_{21}=\mu$ as obtained from WE simulations for the alternative model exhibiting the Allee effect [Eq.~\eqref{alternativeRates}]. In (a) parameters are $N_1 = 490$, $ N_2 = 480$ and $  n_{0,1} = n_{0,2} = 210$, while in (b) $N_1 = 500$, $ N_2 = 70$, $  n_{0,1} = 211 $ and $ n_{0,2} = 70$. }
	\label{fig9sm}
\end{figure}

In the section we briefly present an alternative model which is more realistic in the biological sense, see \textit{e.g.}, Ref.~[20,41]. Here, the Allee effect is locally accounted for by choosing the following birth-death
process:
\begin{eqnarray}\label{alternativeRates}
&&n_i \xrightarrow{B_{n_i}}n_i+1,\;\;\;\;\;\;\;\;\;\;\; n_i \xrightarrow{D_{n_i}}n_i-1\\
&&B_{n_i}= \frac{N_i n_i^2}{n_i^2 + n_{0, i}^2},\;\;\;\;\;\;\;\; D_{n_i}= n_i \nonumber ,
\end{eqnarray}
where $N_i > 0$ is the carrying capacity, $0< n_{0, i} < 1/2$ is the threshold parameter, and migration between patch $i$ and $j$ occurs at a  rate $\mu_{ij}$.
We have simulated this model for two patches using the WE simulations, see previous section. Our results, see Fig.~\ref{fig9sm}, indicate that the analysis done for the simple $2A \leftrightarrow 3A$ and $A\rightarrow 0$ is generic, and holds for other models exhibiting the Allee effect. In particular, our simulations demonstrate the existence of $\mu_{crit}$ and $\mu_{opt}$ for some parameter regimes [Fig.~\ref{fig9sm}(a)], while for other parameter regimes we observe a monotone decreasing MTE as a function of $\mu$ [Fig.~\ref{fig9sm}(b)], similarly as shown Figs.~\ref{fig2sm}(a) and \ref{fig3sm}(a).

\bibliographystyle{apsrev4-1}
\bibliography{VA_bib}

\begin{thebibliography}{47}
\bibitem [{r14()}]{r14}%
  \BibitemOpen
  \href@noop {} {\bibinfo  {journal} {Here, we assume that $r_{4i}=0$ for
  $i=1,2,3$ since the global extinction state (FP4) is absorbing}\
  }\BibitemShut {NoStop}%
\bibitem [{Sco()}]{Scond}%
  \BibitemOpen
  \href@noop {} {\bibinfo  {journal} {In the case where the patches are
  isolated, the MTE is determined by the higher action of the individual
  patches}\ }\BibitemShut {NoStop}%
\end{thebibliography}

\end{document}